\documentclass[fleqn,usenatbib]{mnras}

\usepackage{newtxtext,newtxmath}

\usepackage[T1]{fontenc}

\DeclareRobustCommand{\VAN}[3]{#2}
\let\VANthebibliography\thebibliography
\def\thebibliography{\DeclareRobustCommand{\VAN}[3]{##3}\VANthebibliography}

\usepackage{graphicx}	
\usepackage{amsmath}	
\usepackage{subcaption}
\usepackage{booktabs}
\usepackage{lipsum}
\usepackage[normalem]{ulem}

\newcommand{\uncertainties}[3]{#1{\raisebox{0.5ex}{\tiny$_{-#2}^{+#3}$}}}    

\title[Orbits of cluster galaxies and gas ionization]{Unravelling the orbits of cluster galaxy populations according to their dominant gas ionization source}

\author[G. A. Valk et al.]{
Greique A. Valk,$^{1}$\thanks{E-mail: greique.valk@acad.ufsm.br (GAV)}
Sandro B. Rembold$^{1}$
\\
$^{1}$Departamento de F\'isica, Centro de Ci\^encias Naturais e Exatas, Universidade Federal de Santa Maria, 97105-900, Santa Maria, RS, Brazil}

\date{Accepted 2024 December 16. Received 2024 December 11; in original form 2024 October 28}

\pubyear{2024}

\begin{document}
\label{firstpage}
\pagerange{\pageref{firstpage}--\pageref{lastpage}}
\maketitle

\begin{abstract}
We investigate the kinematical and dynamical properties of cluster galaxy populations classified according to their dominant source of gas ionization, namely: star-forming (SF) galaxies, optical active galactic nuclei (AGN), mixed SF plus AGN ionization (transition objects, T), and quiescent (Q) galaxies. We stack 8\,892 member galaxies from 336 relaxed galaxy clusters to build an ensemble cluster and estimate the observed projected profiles of numerical density and velocity dispersion, $\sigma_P(R)$, of each galaxy population. The MAMPOSSt code and the Jeans equations inversion technique are used to constrain the velocity anisotropy profiles of the galaxy populations in both parametric and non-parametric ways. We find that Q (SF) galaxies display the lowest (highest) typical cluster-centric distances and velocity dispersion values. Transition galaxies are more concentrated and tend to exhibit lower velocity dispersion values than SF galaxies. Galaxies that host an optical AGN are as concentrated as Q galaxies but display velocity dispersion values similar to those of the SF population. MAMPOSSt is able to find equilibrium solutions that successfully recover the observed $\sigma_P(R)$ profile only for the Q, T, and AGN populations. We find that the orbits of all populations are consistent with isotropy in the inner regions, becoming increasingly radial with the distance from the cluster centre. These results suggest that Q galaxies are in equilibrium within their clusters, while SF galaxies have more recently arrived in the cluster environment. Finally, the T and AGN populations appear to be in an intermediate dynamical state between those of the SF and Q populations.

\end{abstract}




\section{Introduction}

    Since the first systematic studies of galaxies, astronomers have noted that such objects exhibit very different morphologies and properties. Regarding morphology, galaxies are commonly split into two broad families: early-type (elliptical/lenticular) and late-type (spiral/irregular) galaxies \citep[e.g.][]{deVaucouleurs_1959_HDP, Sandage_1961_hag}. In addition to the morphological dichotomy, numerous studies have also pointed out that galaxies exhibit colour bimodality, with most galaxies classified as either red or blue \citep[e.g.][]{Strateva_2001_AJ, Baldry_2006_MNRAS, Wyder_2007_ApJS, Salim_2014_SerAJ}. 

    Red galaxies tend to exhibit early-type morphologies; such galaxies also usually host older stellar populations, have lower star formation rates, and are more commonly found in higher-density regions compared to the blue, late-type population \citep[e.g.][]{Strateva_2001_AJ, Balogh_2004_ApJL, Baldry_2006_MNRAS}. Some studies suggest that the fraction of star-forming blue galaxies decreases towards lower redshifts, implying an increase in the fraction of passive red galaxies \citep[e.g.][]{Wolf_2003_A&A, Blanton_2006_ApJ, Willmer_2006_ApJ}. The decline of star formation in galaxies is commonly referred to as \textit{quenching} and occurs when the cold gas ($<100$\,K) within a galaxy, which serves as fuel for star formation, is no longer available. There are several processes that can interrupt the supply of cold gas used to form stars in a galaxy, and their relative efficiencies remain under debate. These processes are generally divided into internal and external ones. Outflows produced by Active Galactic Nuclei (AGN) and supernova feedback are examples of internal processes that can either remove cold gas from galaxies or heat their gas, thereby suppressing star formation \citep[e.g.][]{Springel_2003_MNRAS, Bower_2006_MNRAS, Croton_2006_MNRAS, Agertz_2013_ApJ}. Additionally, star formation can also diminish due to morphological quenching, where the presence of a stellar bulge prevents fragmentation of the gas disk \citep[e.g.][]{Martig_2009_ApJ}. In fact, the colours and morphologies of galaxies also correlate with their gas content: early-type galaxies exhibit lower fractions of cold atomic and molecular gas, as well as warm ionized gas ($10^4$\,K), compared to late-type galaxies \citep[e.g.][]{Roberts_1994_ARA&A, Macchetto_1996_A&AS, Calette_2018_RMxAA}. In contrast to these secular processes, external processes, such as mergers, are associated with interactions between galaxies and their environment \citep[for a review see e.g.][]{Boselli_2006_PASP}. 
    
    The gas present in a galaxy can be ionized by young stars (O and B types) produced in recent star formation events. A significant fraction of AGN hosts exhibit nuclear emission that cannot be explained solely by ionization from stars. These occur due to the accretion of matter into a supermassive black hole located at the galaxy's centre, which produces highly energetic photons that ionize the surrounding gas \citep[e.g.][]{Lynden-Bell_1969_Natur, Rees_1984_ARA&A, Netzer_2015_ARA&A}. Star-forming regions and AGN can coexist in galaxies, and objects where both mechanisms are responsible for ionizing gas are commonly referred to as \emph{transition} or \emph{composite} objects \citep[e.g.][]{Baldwin_1981_PASP, Ho_1993_ApJ, Kauffmann_2003_MNRAS, Kewley_2006_MNRAS}. In addition to galaxies that present ionized gas, either through star formation or AGN activity (or both), there are also objects where the emission lines are very weak or absent, commonly referred to as retired or passive galaxies \citep[see e.g.][]{CidFernandes_2010_MNRAS}. In the former case, the gas is ionized by a population of hot, low-mass evolved stars (HOLMES); in the latter, the galaxies either do not present ionized gas or the ionization is so weak that telescopes cannot detect. 

    It is well-known that galaxy properties vary with their environments. Galaxies residing in high-density regions, such as groups and clusters, tend to exhibit early-type morphologies, while spiral and irregular morphologies are more common in the field. This trend, known as the morphology-density relation, was first identified by \cite{Dressler_1980_ApJ} and has been confirmed by numerous other studies \citep[e.g.][]{Postman_1984_ApJ, Einasto_1987_MNRAS, Treu_2003_ApJ, Huertas-Company_2009_A&A, Houghton_2015_MNRAS}. In addition to the morphological differences, galaxies in high-density regions are redder, have lower star formation rates, and host older stellar populations compared to similar galaxies in low-density regions \citep[e.g.][]{Kauffmann_2004_MNRAS, Weinmann_2006_MNRAS, Weinmann_2009_MNRAS}. Since local density can vary by several orders of magnitude from the central regions of clusters to their outskirts, galaxy clusters serve as excellent natural laboratories for studying galaxy evolution, allowing us to investigate the physical processes responsible for the observed properties of galaxies. To explain the dependence of the galaxy properties on their environments, several mechanisms have been proposed to act in groups and clusters \citep[for a review see e.g.][]{Boselli_2006_PASP}. The high-density regions of groups and clusters are conducive to interactions between galaxies. Such interactions can give rise to mergers, producing significant morphological transformations in the galaxies involved \citep[e.g.][]{Gerhard_1981_MNRAS}. Moreover, mergers can cause the gas to lose angular momentum and infall towards the central regions, inducing starbursts and/or triggering AGN activity \citep[e.g.][]{Noguchi_1988_A&A, Hernquist_1995_ApJ, Mihos_1996_ApJ, Gao_2020_A&A}. Mergers are more common in groups or at the outskirts of rich clusters because they require lower relative velocities between the galaxies involved. On the other hand, the cumulative effect of close encounters between galaxies in clusters, referred to as galaxy harassment, can also affect the structure of the galaxy and cause gas to fall towards the central regions \citep[e.g.][]{Moore_1998_ApJ}. Galaxies moving through the intra-cluster medium (ICM) are subject to both dynamical friction \citep[e.g.][]{Chandrasekhar_1943_ApJ} and ram-pressure stripping \citep[e.g.][]{Gunn_1972_ApJ}. Dynamical friction, which is more effective for more massive objects, causes galaxies to lose energy and angular momentum, driving them towards the central regions. This process is commonly invoked to help explain the properties exhibited by central galaxies in clusters \citep[e.g.][]{Martizzi_2012_MNRAS}. Ram-pressure stripping, which occurs due to the pressure exerted by the hot, X-ray emitting gas ($10^7 - 10^8$\,K) that permeates the ICM into the gas component of galaxies, can directly affect its star formation \citep[e.g.][]{Farouki_1980_ApJ, Abadi_1999_MNRAS, Tonnesen_2009_ApJ, Boselli_2022_A&ARv, Wright_2022_MNRAS}. In addition, ram-pressure and tidal forces can also remove the halo of hot gas that surrounds the galaxies and serves as a fuel reservoir for the replenishment of star formation in these objects, resulting in lower star formation rates in galaxies at high-density regions compared to similar ones in the field \citep[e.g.][]{Balogh_1999_ApJ, Balogh_2000_ApJ, vandenBosch_2008_MNRAS}.

    The time-scales and spatial regions in which these environmental mechanisms operate are different \citep[see e.g.][]{Boselli_2006_PASP}. Furthermore, the efficiency of some of these processes correlates with the orbital characteristics of the galaxies. Ram-pressure stripping, starvation, and dynamical friction, for example, are more efficient in galaxies that are accreted with more radial orbits into the cluster environment \citep[e.g.][]{Vollmer_2001_ApJ, Jaffe_2018_MNRAS, Lotz_2019_MNRAS}. Consequently, the orbits of galaxies within clusters provide a powerful tool for assessing the relative efficiency of these mechanisms in altering galaxy properties. Galaxies falling towards clusters through filaments are expected to have more radial orbits and, since the cluster environmental mechanisms have not yet had sufficient time to affect these objects, their properties will still be determined by their original environments, be they field galaxies or pre-processed galaxies in low-mass haloes along the filaments themselves \citep[e.g.][]{Martinez_2016_MNRAS, Salerno_2019_MNRAS}. In contrast, galaxies that have resided longer within the cluster environment have already been affected by these various mechanisms, which simultaneously isotropize their orbits, remove their gas, and change their morphologies. 

    In contrast to the comparatively well-understood relation between quenching of star formation and environment, the connection between AGN activity and environment is not yet so clear. Numerous studies have shown that the frequency of optical nuclear activity is lower in high-density regions \citep[e.g.][]{Gavazzi_2011_A&A, Sabater_2013_MNRAS, Lopes_2017_MNRAS, RodriguezDelPino_2023_A&A}, with some findings indicating that the AGN fraction decreases towards the centre of clusters \citep[e.g.][]{Gavazzi_2011_A&A,Pimbblet_2013_MNRAS, Lopes_2017_MNRAS}. However, some studies suggest that this relationship with the environment depends on galaxy type. According to \cite{vonderLinden_2010_MNRAS}, the fraction of star-forming galaxies that host a powerful optical AGN ($L[\text{OIII}] > 10^7 \, L_\odot$) is independent of the cluster-centric distance, while the fraction of quiescent galaxies hosting a weak AGN decreases as one approaches the cluster centre. Similar results indicate that the fraction of late-type/blue galaxies hosting an AGN shows little dependence on the environment, whereas the fraction of early-type/red galaxies is higher in low-density regions \citep[e.g.][]{Hwang_2012_A&A, Sohn_2013_ApJ, Miraghaei_2020_AJ}. Conversely to these studies, there are also works that find little dependence between nuclear activity and the environment of their host galaxies \citep[e.g.][]{Slavcheva-Mihova_2011_A&A, Man_2019_MNRAS}.
    
    In addition to the difference in AGN frequency between low- and high-density regions, some works also suggest that AGN activity is triggered by environmental effects. \cite{Rembold_2024_MNRAS} compared the large- and small-scale environments of optical AGNs with a control sample of non-active galaxies matched by redshift, stellar mass and morphology. Their results indicate no difference between the large- and small-scale environments of these two samples. However, the AGN sample was shown to exhibit an excess of non-circular gas motions, which is linked not to AGN feedback, but to the larger tidal fields experienced by AGN hosts compared to non-active galaxies. This suggests that AGN activity is triggered by tidal interactions between AGN hosts and nearby galaxies. A similar result was obtained by \cite{Sabater_2013_MNRAS}, which identified that the AGN fraction (at fixed stellar mass) is enhanced by galaxy interactions. Furthermore, according to \cite{Poggianti_2017_Natur} and \cite{Peluso_2022_ApJ}, galaxies that experience ram pressure are more likely to host an AGN than galaxies that are not ram-pressure stripped.

    These results suggest that environmental mechanisms can affect not only the star formation of a galaxy but also its ability to display optical nuclear activity. Therefore, since the efficiency of such mechanisms depends not only on the galaxy's location within the cluster but also on its orbital properties, orbital analysis can provide valuable insights to improve our understanding of this subject. In this sense, much work has been done over the years to investigate how the orbits of galaxies relate to their properties. Some studies have shown that, at low-redshift ($z\sim 0.1$), passive/early-type/red galaxies exhibit more isotropic orbits compared to star-forming/late-type/blue galaxies \citep[e.g.][]{Mahdavi_1999_ApJ, Biviano_2004_A&A, Munari_2014_A&A, Mamon_2019_A&A}. These differences seem to disappear at high-redshift ($z > 0.4$), with both families being characterized by similar velocity anisotropy profiles, which indicate isotropic orbits in the inner regions that become more radial with increasing cluster-centric distances \citep[e.g.][]{Biviano_2009_A&A, Biviano_2013_A&A, Biviano_2016_A&A}. Therefore, these results suggest an evolution in the orbital profile of the passive/early-type/red galaxies towards lower redshifts, while the profile of star-forming/late-type/blue galaxies remains unchanged. However, there are also works that have found opposite results. \cite{Aguerri_2017_MNRAS}, analysing the orbits of galaxies in the cluster Abell 85, found that red galaxies move on more radial orbits than blue galaxies. The same result was obtained by \cite{Mercurio_2021_A&A}, studying the cluster Abel S1063. Theoretical approaches have also presented divergent results. \cite{Lotz_2019_MNRAS}, using the set of cosmological hydrodynamical simulations \emph{Magneticum Pathfinder}, analysed the orbits of galaxies across a wide range of cluster masses $[(1-90) \times 10^{14} \, M_\odot]$ and redshifts $(0<z<2)$. These authors found that, independent of cluster mass and redshift, star-forming satellite galaxies move on more radial orbits than quiescent satellite galaxies. On the other hand, \cite{Iannuzzi_2012_MNRAS} and \cite{AguirreTagliaferro_2021_A&A}, both employing semi-analytical models applied to the Millennium Simulation \citep{Springel_2005_Natur}, obtained that red galaxies are characterized by more radial orbits compared to blue galaxies.

    Consequently, the relationship between orbital profiles and the properties of galaxy populations remains a subject of ongoing debate.
    In addition, previous studies commonly focus on the orbital profile of passive/early-type/red versus star-forming/late-type/blue galaxies. To the best of our knowledge, no studies have examined the orbital profile of galaxies by splitting them into populations based on the main mechanism responsible for gas ionization. However, the previously mentioned results hint at some relationship between environmental mechanisms and optical nuclear activity. There are processes that can directly trigger an AGN, such as mergers and galaxy harassment \citep[e.g.][]{Sanders_1988_ApJ, Hernquist_1995_ApJ}, while others indirectly affect the AGN by removing the gas reservoir from galaxies, like ram-pressure stripping and starvation \citep[e.g.][]{Poggianti_2017_Natur, Peluso_2022_ApJ}. Therefore, analysing the orbital properties of AGN hosts can provide interesting insights into how the nuclear activity can be triggered or suppressed within galaxy clusters through the action of environmental mechanisms.

    In this paper, we investigate the orbital properties of galaxies in clusters according to the dominant gas ionization agent. In particular we determine, for the first time,  the velocity anisotropy profiles of AGN and transition galaxies and compare them with those of the star-forming and quiescent galaxies. Our main goal is to derive and compare the kinematical and dynamical properties of these galaxy populations in order to improve the understanding of how the environment can modify galaxy properties. 
    The paper is organized as follows. In Section~\ref{sec:Data} we present the data sample. The methods used to extract the orbital parameters of our galaxy populations are described in Section~\ref{sec:Methods}. In Section~\ref{sec:Gaussian_vs_Non-Gaussian_systems} we investigate the biases introduced by including unrelaxed systems in the dynamical analysis and present an approach to mitigate this issue. In Section~\ref{sec:Galaxy_populations} we present the results of the orbital analysis of galaxies according to their dominant gas ionization mechanism. The results are discussed in Section~\ref{sec:Discussion} and our main conclusions are summarized in Section~\ref{sec:Conclusions}. Throughout this paper we adopt $H_0 = 67.8$ km\,s$^{-1}$\,Mpc$^{-1}$, $\Omega_m = 0.308$ and $\Omega_\Lambda = 0.692$ \citep{PlanckCollaboration_2016_A&A}.
    

\section{Data}
\label{sec:Data}

    In this study, we use the galaxy and group catalogues presented by \citet[][hereafter T17]{Tempel_2017_A&A}. These catalogues are based on the Sloan Digital Sky Survey \citep[SDSS,][]{York_2000_AJ} data release 12 \citep{Eisenstein_2011_AJ, Alam_2015_ApJS} and are complemented by 1\,349 redshifts from other surveys. The final galaxy sample comprises 584\,449 objects with Petrosian $r$-band magnitudes brighter than 17.77 and redshifts below $z = 0.2$. The galaxies were associated to groups using a group finder algorithm based on the friends-of-friends (FoF) method \citep[e.g.][]{Turner_1976_ApJS, Beers_1982_ApJ, Zeldovich_1982_Natur}. Additionally, a membership refinement algorithm was applied to improve the reliability of the groups identified by the FoF method. T17 computed a number of parameters for each group, and those of particular interest in this work are the velocity dispersion and the virial radius of the group. The velocity dispersion of a group was computed from the variance of the line-of-sight velocities of its confirmed members. The virial radius of each system is uniquely defined by its virial mass, which was computed using the virial theorem and depends exclusively on the group extent in the sky, the velocity dispersion, and the assumed mass density profile. The virial masses in T17 are computed assuming a NFW profile \citep{Navarro_1996_ApJ} and the mass-concentration relation of \cite{Maccio_2008_MNRAS}. The final group catalogue comprises 88\,662 systems with at least two members, among which 498 merging systems were identified due to overlaps between the virial radii of multiple clusters.

    To conduct a reliable dynamical analysis, a minimum number of members per system is required. This ensures confidence in the physical parameters derived for the clusters and guarantees that the estimated locations of their centres are meaningful. For this reason, we select galaxies from the T17 galaxy and group catalogues that belong to systems with 20 or more members, resulting in 642 clusters and 23\,977 galaxies. 

    The fluxes and equivalent widths ($W$) of the emission lines H$\alpha$, H$\beta$, [NII]$\lambda\,6584$ and [OIII]$\lambda\,5007$ for the galaxies in our sample are downloaded from the SDSS Catalogue Archive Server\footnote{\url{https://skyserver.sdss.org/casjobs/}} (CAS). The measurements of the emission lines were carried out by \citet{Thomas_2013_MNRAS}, using adaptations of the codes Gas AND Absorption Line Fitting \citep[GANDALF,][]{Sarzi_2006_MNRAS} and penalized PiXel Fitting \citep[pPXF,][]{Cappellari_2004_PASP}, and are available in the SDSS database (table \emph{emissionLinesPort}). The stellar population models from \citet{Maraston_2011_MNRAS} and \citet{Thomas_2011_MNRAS} were used for the continuum.
 

\section{Methods}
\label{sec:Methods}


\subsection{Ensemble Cluster}
\label{subsec:Ensemble_Cluster_Methods}

Dynamical analyses of galaxy clusters require a minimum number of galaxies per system to provide reliable results with low uncertainties. However, this criterion is generally not satisfied by individual clusters due to the limitations in spectroscopic surveys. A common approach to circumvent this limitation is to stack the clusters to create a `pseudo-cluster', referred to as an \textit{ensemble cluster}, and analyse it under the assumption that it is representative of each cluster in the sample \citep[e.g.][]{Katgert_2004_ApJ, Biviano_2016_A&A, Mamon_2019_A&A, Biviano_2021_A&A}. This method is anchored by numerical simulations which predict a global mass profile, $M(r)$, for haloes of dark matter \citep[e.g.][]{Navarro_1996_ApJ} and by the fact that the $M(r)$ profile is weakly dependent on halo mass and redshift \citep[e.g.][]{DeBoni_2013_MNRAS,Biviano_2016_A&A}. Finally, this method is typically applied only to clusters without substructures, as the existence of a fundamental plane relating certain global properties of clusters suggests that these systems form a homologous set \citep[e.g.][]{Adami_1998_A&A_1}. In contrast, deviations from the fundamental plane by clusters with substructures indicate that these systems violate the homology assumption \citep[e.g.][]{Beisbart_2001_A&A}.

In order to perform the dynamical analysis of the clusters in our sample we stack them to create an ensemble cluster. Let $R_i$ and $v_i$, respectively, be the projected cluster-centric distance and the line-of-sight (hereafter LOS) velocity of the $i$-th galaxy, and $r_{200,j}$ and $\sigma_{v,j}$ be the virial radius\footnote{Throughout this paper, we refer to the `virial radius' as the radius of a sphere in which the mean matter density is 200 times higher than the mean density of the Universe, represented by $r_{200}$.} and the velocity dispersion, respectively, of its parent ($j$-th) cluster. Then, $R_{i,j} = (R_i /r_{200,j}) \langle r_{200} \rangle $ and $v_{i,j} = (v_i / \sigma_{v,j}) \langle \sigma_v \rangle$ are, respectively, the projected cluster-centric distance and the LOS velocity of the $i$-th galaxy in the ensemble cluster. Here, $\langle r_{200} \rangle$ and $\langle \sigma_v \rangle$ are the average values of $r_{200}$ and $\sigma_v$ of the clusters that constitute the ensemble cluster, respectively. Similarly to $\langle r_{200} \rangle$ and $\langle \sigma_v \rangle$, we also compute the mean redshift of the ensemble cluster, $\langle z \rangle$. Creating the ensemble cluster following the above procedure allows us to obtain $R_{i,j}$ and $v_{i,j}$ for each galaxy in units of distance and velocity, respectively. This means that we can analyse all the galaxies as being part of a single (pseudo-)cluster with the $r_{200}$ and $\sigma_{v}$ typical of the sample. Furthermore, this method also preserves the relative positions and velocities of the galaxies in their parent clusters within the ensemble cluster.


\subsection{Observed projected profiles}
    \label{subsec:Observed_projected_profiles_Methods}

    The galaxy numerical density and velocity dispersion profiles are the fundamental observables from which galaxy clusters' anisotropy profiles are drawn. In the following, we explain how we estimate these profiles for the ensemble cluster.
        
    The projected numerical density, $I$, is estimated in circular rings concentric with the centre of the ensemble cluster. For a circular ring with radius $R$ and thickness $dR$, the projected numerical density is defined as $I(R+dR/2) = N/A$, where $N$ is the number of galaxies within the ring and $A$ is the area of the ring. We estimate $I$ in radial bins with a fixed number of galaxies, except for the last bin, to create a radial profile of $I(R)$. We then fit\footnote{All the fitting procedures realized in this work were performed using the curve\_fit function, which employs the Levenberg-Marquardt technique, from the \textsc{Python} Scipy package, available at \url{https://scipy.org/}.} to the observed number density profile an analytical profile given by the Abel integral

    \begin{equation}\label{eq:Abel_Integral}
        I(R) = 2 \int_{R}^{\infty} \frac{r \nu(r) dr}{\sqrt{r^2-R^2}},
    \end{equation}
    
    \noindent where the numerical density profile, $\nu(r)$, is the NFW profile, defined as
    \begin{equation}\label{eq:NFW_Profile}
        \nu(r) = \frac{\nu_0}{r(r+r_\nu)^2},
    \end{equation}

    \noindent with $r_\nu$ and $\nu_0$ being the scale radius and the normalization factor of the NFW profile, respectively. To perform the fit, we fix the upper limit in the integral to $60\,000$\,kpc. The choice of this limit value is not relevant for the results, provided that the chosen value is sufficiently far from the fitting region. 

    The LOS velocity dispersion, $\sigma_P$, is estimated using
    \begin{equation}\label{eq:sigmaP_percentile}
        \sigma_P = \dfrac{\text{IQR}}{1.349},
    \end{equation}

    \noindent where $\text{IQR} = p_{75}(v)-p_{25}(v)$ is the interquartile range, and $p_x$ denotes the $x$-th percentile of the velocity distribution. We chose to use the IQR to compute $\sigma_P$ to avoid potential issues at the edges of the velocity distribution, due to the fact that the FoF method imposes strict boundaries at the edges of the distribution of peculiar velocities during the process of identifying galaxy clusters. Furthermore, the removal of galaxies during the membership refinement process will also affect the velocity distribution. Similarly to the $I(R)$ profile, we estimate the velocity dispersion in radial bins with a fixed number of galaxies (except the last bin) to create a radial profile of $\sigma_P(R)$. The uncertainties in each bin of $\sigma_P(R)$ are computed by bootstraps with 1\,000 resamples. 

    The inversion of the Jeans equations (Section~\ref{sec:Inversion_of_the_Jeans_Equations_Methods}) requires a continuous version of the $\sigma_P(R)$ profile. We have obtained the $\sigma_P(R)$ profiles for a number of NFW density/anisotropy profiles pairs with varying parameters, and have found that the functional form 
    \begin{equation}\label{eq:fit_sigmaP}
        \sigma_P(R) = \frac{aR^d}{1+bR^e\exp(cR)},
    \end{equation}
    
    \noindent was successful in describing the shape of the projected velocity dispersion with good accuracy. We use this expression to fit our observed, binned velocity dispersion profiles. We have also found that this method results in more robust estimates of the overall velocity profile shape than a simple smoothing kernel on the binned velocity dispersion profile. 
    


\subsection{MAMPOSSt}
\label{subsec:MAMPOSSt_Methods}

   
    We have investigated the dynamical properties of the ensemble cluster using the MAMPOSSt (Modelling Anisotropy and Mass Profiles of Observed Spherical Systems) code, developed by \citet{Mamon_2013_MNRAS}. The code performs a maximum likelihood fit of the tracers (galaxies) in the projected phase-space (PPS), using parametric forms for the $\nu(r)$, $M(r)$ and $\beta(r)$ profiles. Here, $\beta(r)$ denotes the velocity anisotropy profile, defined as 
    \begin{equation}
        \beta(r) = 1 - \frac{\sigma_\theta^2(r)+\sigma_\phi^2(r)}{2\sigma_r^2(r)},
    \end{equation}

    \noindent where $\sigma_\theta$ and $\sigma_\phi$ are the two tangential components of the velocity dispersion, and $\sigma_r$ is the radial component. In spherical symmetry, we must have $\sigma_\theta = \sigma_\phi$. Isotropic orbits correspond to $\beta = 0$, while purely radial and tangential orbits are characterized by $\beta = 1$ and $\beta = -\infty $, respectively. Finally, MAMPOSSt requires only the cluster-centric distances and velocities of the galaxies as input without any kind of binning in the observed data.
    
    MAMPOSSt is based on the Jeans equations. This implies that dynamical equilibrium is assumed for the galaxies being analysed and that the solutions provided are equilibrium solutions. To satisfy this condition, a common approach consists in restricting the dynamical analysis only to galaxies with projected cluster-centric distances ($R$) smaller than the virial radius of the system \citep[e.g.][]{Biviano_2016_A&A, Biviano_2021_A&A}. However, it is important to mention that this approach does not fully meet the equilibrium condition. Although we expect the galaxies lying inside the virial sphere to be closer to equilibrium, real samples suffer from projection effects and contamination from interlopers. Despite this issue, we decided to follow the same methodology employed by other authors in the literature and apply MAMPOSSt only to galaxies with $R \leq r_{200}$, removing all objects with $R > r_{200}$ from our sample. Also, the dynamics of close satellites of a cluster's central galaxy (or Brightest Cluster Galaxy, BCG) is driven mostly by the gravitational potential of the central galaxy instead of the overall cluster potential. We have therefore removed from the sample all galaxies closer than $0.1 r_{200}$ to its parent cluster centre. This removal also helps to reduce the impact of fibre collisions on the projected number density profiles. Finally, MAMPOSSt assumes both spherical symmetry ($\sigma_\theta = \sigma_\phi$) for the ensemble cluster -- which is not an issue because our ensemble cluster is, by construction, spherically symmetric -- and a Gaussian 3D velocity distribution.
    
    The execution of MAMPOSSt requires the user to specify the functional forms of the number density, total mass density and velocity anisotropy profiles. For the numerical density profile, we chose a NFW profile (equation~\ref{eq:NFW_Profile}). As shown in Section~\ref{subsec:Recentralization} and Section~\ref{subsec:I_profiles_Pops}, the NFW profile for $\nu(r)$, after being projected by equation~(\ref{eq:Abel_Integral}), fits the $I(R)$ profiles of the galaxy populations in our sample very well. MAMPOSSt allows constraints on the parameters, which can be used to reduce the number of free parameters in the fit. We chose to execute MAMPOSSt in the so-called \textit{Split} mode \citep[see][]{Mamon_2013_MNRAS}. In this mode, we externally fit the $\nu(r)$ profile and provide to MAMPOSSt only the best-fitting value for $r_\nu$, with its respective uncertainty. In this mode, MAMPOSSt restricts the maximum likelihood fit to the velocity space.

    We also adopt a NFW profile for the cluster mass density, $\rho(r)$. Therefore, the mass profile is given by
    \begin{equation}\label{eq:Mass_profile}
        M(r) = M_{200} \frac{\ln(1+r/r_{-2})-r/r_{-2}(1+r/r_{-2})^{-1}}{\ln(1+c)-c/(1+c)},
    \end{equation}
    
    \noindent where $r_{-2}$ is the radius where the logarithmic slope of the mass density profile is equal to $-2$, $c\equiv r_{200}/r_{-2}$ represents the concentration of the mass profile, and $M_{200}$ is the mass contained within $r_{200}$. The $M_{200}$ value is obtained from $r_{200}$ through the equation
    \begin{equation}
        M_{200} = 200\frac{H^2(z)r^3_{200}}{2G},
    \end{equation}
    
    \noindent where $H(z)$ is the Hubble parameter. Therefore, the mass profile is completely defined by the $r_{200}$ and $r_{-2}$ parameters.

    The MAMPOSSt estimation of the mass profile allows for the contribution of three components: dark matter, massive tracers, and a central black hole. In the remainder of this work, we consider the galaxies as being tracers without mass of the cluster potential. We also neglect any contribution from a possible central black hole. Therefore, the mass profile determined by MAMPOSSt takes into account only the dark matter contribution. 

    To describe the velocity anisotropy profile, we adopt the generalized Osipkov-Mettitt profile \citep{Osipkov_1979_SvAL, Merritt_1985_MNRAS}
    \begin{equation}\label{eq:beta}
        \beta(r) = \beta_0+(\beta_\infty-\beta_0)\frac{r^2}{r^2+r_\beta^2}.
    \end{equation}
    \noindent The $\beta_0$ and $\beta_\infty$ parameters represent the velocity anisotropy at $r=0$ and $r=\infty$, respectively. Additionally, the scale radius of the profile, $r_\beta$, controls the radius where $\beta$ transitions from the inner to the outer value. This $\beta(r)$ profile is characterized by three free parameters, making it quite flexible. As a result, the number of free parameters in the MAMPOSSt execution can be high if we consider several tracer populations. Therefore, we execute MAMPOSSt with the constraint $r_\nu = r_\beta$, \textit{when necessary}, to reduce the number of free parameters. This constraint on the MAMPOSSt execution, already considered in the literature \citep[e.g.][]{Mamon_2019_A&A, Biviano_2024_ApJ}, is motivated by results extracted from numerical simulations \citep{Mamon_2010_A&A} and is referred to as \textit{TAND} \citep[Tied Anisotropy Number Density, see][]{Mamon_2019_A&A}.


\subsection{Inversion of the Jeans Equations}
\label{sec:Inversion_of_the_Jeans_Equations_Methods}

    Dynamical analyses based on the Jeans equations are susceptible to degeneracy between the $M(r)$ and $\beta(r)$ profiles. This so-called mass-anisotropy degeneracy arises because there are only two equations to solve for three unknown profiles ($M$, $\sigma_r$ and $\beta$). MAMPOSSt uses the full velocity distribution of the galaxies to break this degeneracy, while methods based on the inversion of the Jeans equations require knowledge about either $M(r)$ or $\beta(r)$. The mass profile of galaxy clusters is generally well represented by a NFW model \citep[e.g.][]{Geller_1999_ApJL, Biviano_2003_ApJ, Oguri_2012_MNRAS, Biviano_2013_A&A, Okabe_2013_ApJL}. However, cluster-sized haloes extracted from simulations tend to present very diverse $\beta(r)$ profiles \citep[e.g.][]{Mamon_2013_MNRAS}.

    This renders the shape of the $\beta(r)$ profile the most uncertain of the MAMPOSSt outputs due to the constraints imposed by a single functional form for $\beta(r)$ -- in particular with the constraint of the \textit{TAND} mode. Due to this, we perform the inversion of the Jeans equations (hereafter IJE), following the method of \citet{Solanes_1990_A&A}, to get a non-parametric estimate of the $\beta(r)$ profile. The IJE approach requires the $M(r)$ profile of the ensemble cluster, which we adopt as the $M(r)$ profile estimated by MAMPOSSt.

    The IJE approach uses the $\sigma_P(R)$, $I(R)$ and $M(r)$ profiles to estimate the $\beta(r)$ profile. Unlike MAMPOSSt, IJE requires continuous versions of the $\sigma_P(R)$ and $I(R)$ profiles, as well as their extrapolations to high radial distances. Instead of estimating a unique $\beta(r)$ profile from the observed $\sigma_P(R)$ and $I(R)$ profiles and the $M(r)$ profile from MAMPOSSt, we adopt a bootstrap approach to estimate an average $\beta(r)$ profile and its respective uncertainties. In short, we draw random subsamples from the original galaxy sample and derive the $\sigma_P(R)$ profile for each subsample. We then combine the velocity dispersion profile of each subsample with randomly chosen values for the parameters defining the $M(r)$ and $\nu(r)$ profiles across the parameter space in the MAMPOSSt solution. By applying the IJE method, we obtain a single estimate of the $\beta(r)$ profile per subsample. Finally, we determine the median $\beta(r)$ profile and its uncertainties across 1\,000 realizations. 

    \subsection{Identifying the clusters' central positions}
    \label{subsec:Identifying_the_clusters_central_positions}
    
    The process of combining galaxies from several clusters to create an ensemble cluster requires accurate estimation of the central coordinates for all the clusters involved. Clusters with poorly defined centres can introduce errors in the estimates of the $I(R)$ and $\sigma_P(R)$ profiles, thereby affecting the derived anisotropy profiles. Based on this, before carrying out the dynamical analysis, we decided to check if the centre of the clusters as defined by T17 correctly maps the density peak of the cluster. To achieve this, we compare the $I(R)$ profiles and their respective NFW fitted profiles for ensemble clusters that differ only in the centre position considered for each individual cluster during the process of creating the ensemble clusters. Firstly, we consider the centre as estimated by T17, which we call the observed (O) centre. Then, we re-estimate the centre of each individual cluster using the peak of a gaussian kernel applied to the 2D distribution of galaxies; we refer to this new location as the corrected (C) centre. Note that we do not consider the centre position defined by the location of the BCG, as we only have information of the most luminous (rank 1) galaxy in each system. As discussed in Section~\ref{subsec:Presence_of_central_galaxies}, these galaxies are not always equivalent to BCGs.

    We verify that the $I(R)$ profile obtained considering the O centre tends to be flatter in the central region and decreases more slowly in the external region than the profile obtained considering the C centre. Additionally, the NFW profile provides a better fit for the profiles estimated considering the C centre. We illustrate this improvement in Section~\ref{subsec:Recentralization}. Based on these results, we argue that the C centre position better maps the radial density profiles of cluster galaxies in our sample. We therefore adopt the corrected centre position in the remainder of this work, rather than the one obtained by T17, to construct the ensemble clusters. It is important to note that we re-estimate the centre position but keep the $r_{200}$ value as estimated by T17. 
        
    \subsection{Characterizing the gas ionization source}
    \label{subsec:Identification_of_the_gas_ionization_source}
    
    The galaxies in our sample are classified into populations based on the main source of gas ionization, namely: star-forming (SF), optical active galactic nuclei (AGN), transition objects (T, SF+AGN ionization), and quiescent (Q). The classification is performed using the BPT-NII \citep{Baldwin_1981_PASP} and WHAN \citep{CidFernandes_2011_MNRAS} optical diagnostic diagrams simultaneously. Galaxies below the \citet{Kauffmann_2003_MNRAS} line in the BPT diagram and with \mbox{$W_{\mathrm{H}\alpha} > 3\,$ \AA\,} are classified as SF. Transition galaxies are located between the \citet{Kauffmann_2003_MNRAS} and \citet{Kewley_2001_ApJ} lines in the BPT diagram and have \mbox{$W_{\mathrm{H}\alpha} > 3\,$ \AA}. AGN galaxies are located above the \citet{Kewley_2001_ApJ} line and have \mbox{$W_{\mathrm{H}\alpha} > 3\,$  \AA\,} and \mbox{$\log(\mathrm{[N II]}/\mathrm{H}\alpha) > -0.4$}. Finally, quiescent galaxies have \mbox{$W_{\mathrm{H}\alpha} < 3\,$ \AA}, regardless of their position on the BPT diagram. 

    The requirement of four emission lines to use the BPT and WHAN optical diagnostic diagrams prevented the classification for a considerable fraction of galaxies, due to the absence of information for at least one of the emission lines required by those. This problem arises due to a numerous sample of galaxies with weak emission lines, generally present in spectroscopic samples \citep[see e.g.][]{CidFernandes_2010_MNRAS}. Several of these galaxies are quiescent galaxies with intrinsically weak lines. Therefore, we choose to use the information from the available lines to try to classify them as quiescent galaxies.
    
    Galaxies where the H$\alpha$ line is detected can be classified as quiescent if $W_{\mathrm{H}\alpha} < 3 $ \AA, without considering the line ratios on the other axes of the optical diagnostic diagrams. In contrast, if the H$\alpha$ line is not available, then the galaxy cannot be classified using the optical diagnostic diagrams. To avoid the loss of these galaxies, we analyse the behaviour of the remaining emission lines. We compare the distribution of $W_{\mathrm{[NII]}}$ for galaxies classified as quiescent and non-quiescent (SF+AGN+T) using the optical diagnostic diagrams and find that these differ substantially. Quiescent galaxies tend to present significantly lower values of $W_{\mathrm{[NII]}}$ than non-quiescent galaxies. Therefore, in the absence of the H$\alpha$ line, we classify as quiescent the galaxies with $W_{\mathrm{[NII]}}$ below the 5th percentile of the distribution of $W_{\mathrm{[NII]}}$ of the non-quiescent population.


    As discussed earlier, both MAMPOSSt and IJE assume that the systems under analysis are in dynamical equilibrium. In our sample, in addition to the merging clusters identified by T17, there is some degree of contamination from systems that are far from equilibrium. Therefore, it is necessary to assess the impact of contamination from this type of system and to develop tools to minimize their influence as much as possible. In the next section, we address this issue.
    

\section{Testing the effects of cluster non-equilibrium on the derived anisotropy parameters}
\label{sec:Gaussian_vs_Non-Gaussian_systems}

In this section, we investigate the derived mass and anisotropy profiles of clusters of galaxies far from the dynamical equilibrium when applying the methodology presented in the previous section. Our main goal is to obtain the specific signature of this kind of population and to evaluate how severely the inclusion of such systems in our ensemble cluster contaminates the anisotropy profile derived by means of MAMPOSSt and IJE.


\subsection{Gaussianity Classification}
\label{subsec:Gaussianity_Classification}


In order to perform this analysis, we separate the clusters of our sample into two sub-samples: relaxed and non-relaxed clusters. To make an absolute inference on the degree of dynamical evolution of clusters of galaxies is not an easy task due to the large variability on the typical observables associated to non-equilibrium \citep[e.g.][]{Einasto_2012_A&A, Lopes_2018_MNRAS}. For our purposes in this section, however, we only need to classify clusters according to an objective dynamical state inference tool. A simple methodology that has been applied to investigate the dynamical state of clusters of galaxies is based on the Hellinger Distance (HD) estimator. Introduced by \cite{Ribeiro_2013_MNRAS} and extensively tested in \cite{deCarvalho_2017_AJ}, this method allows one to separate clusters into relaxed and unrelaxed ones according to the shape of their LOS galaxy velocity distribution: relaxed (unrelaxed) clusters present velocity distributions that are (are not) compatible with a gaussian distribution. We therefore refer to relaxed and unrelaxed clusters in this section as gaussian (G) and non-gaussian (NG) clusters.

The $\rm{HD}$ statistics estimates how close two probability distributions, $p(x)$ and $q(x)$, are to each other and is computed by
\begin{equation}
    \rm{HD^2(p,q)} = \frac{1}{2} \int \left( \sqrt{p(x)}-\sqrt{q(x)} \right)^2 dx.
    \label{eq:hellinger_distance}
\end{equation}

\noindent In our case, we are interested in estimating how close the observed LOS velocity probability distribution, $q(x)$, is to a Gaussian one, $p(x)$.

In this work, we calculate the $\rm{HD}$ and classify clusters into G/NG systems using a methodology similar to that applied by \cite{deCarvalho_2017_AJ}. We first generate $\rm{HD}$ distributions as a function of the number $N$ of elements in the sample by resampling a strictly Gaussian distribution with size $N$. For each $N$, we define a threshold below which $95.45\%$ of the $\rm{HD}$ values\footnote{This is the percentile that was found to be closest to the $\rm{HD}_{\rm{Median}}+3\sigma_{\rm{HD}}$ threshold proposed by \cite{deCarvalho_2017_AJ}.} are located. Then, for a given cluster containing $N$ galaxies, we perform 10\,000 random resamplings of the LOS velocities of the galaxies, obtain the $\rm{HD}$ for each realization, and calculate the fraction $\rm{HD}_R$ of the realizations that resulted in a larger $\rm{HD}$ than the threshold value for $N$ objects. If $\rm{HD}_R > 0.5$, the cluster is classified as Non-Gaussian (NG) with a reliability value of $100 \rm{HD}_R\%$, while if $\rm{HD}_R \leq 0.5$ the cluster is classified as Gaussian (G) with a reliability value of $100(1-\rm{HD}_{R}) \%$.

\begin{figure*}
    \centering
    \includegraphics[width=\linewidth]{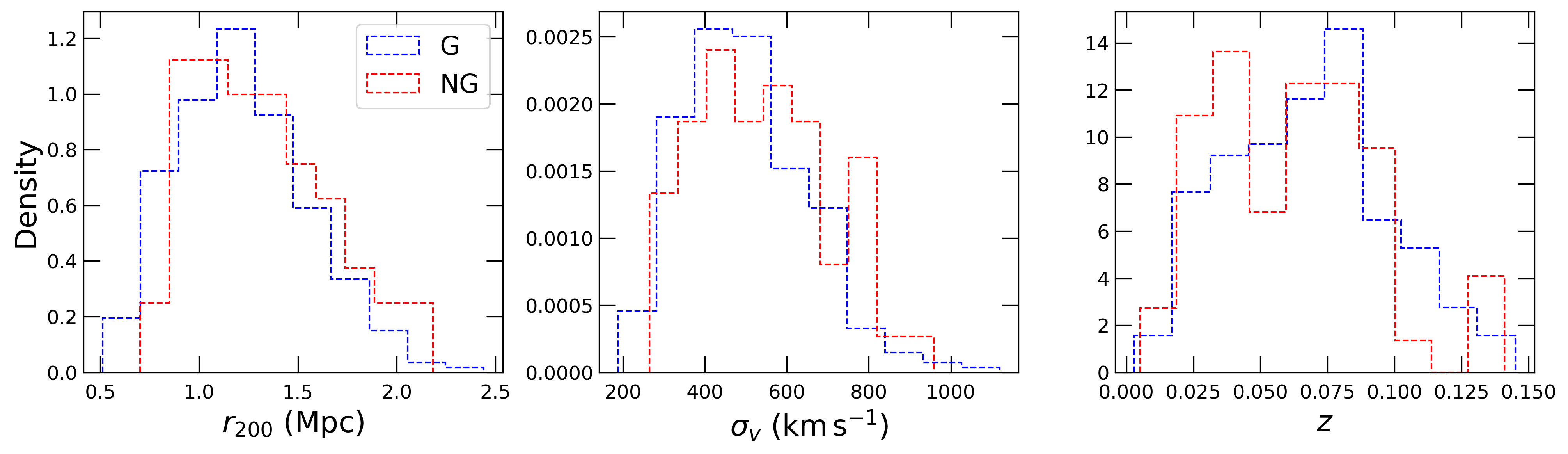}
    \caption{Distributions of virial radius $r_{200}$ (left panel), LOS velocity dispersion $\sigma_v$ (central panel), and redshift $z$ (right panel) values for G (dashed blue line) and NG (dashed red line) clusters.}
    \label{fig:distribution_r200_sigma_z}
\end{figure*}

We apply this method for the 642 clusters in our sample, and obtain 529 G ($\sim 82\%$) and 113 NG ($\sim 18\%$) clusters. These fractions are in agreement with those obtained by \citet{deCarvalho_2017_AJ}, which are $76 \%$ and $24 \%$ for G and NG clusters, respectively. In this step, we decided to keep the classification as NG systems only for those clusters that present a reliability value equal to or higher than $70\%$. This guarantees that we assign as NG systems only clusters whose LOS velocity distribution is far enough from a Gaussian distribution. Applying this criterion, we obtained 588 G ($\sim 92\%$) systems with 21\,633 galaxies and 54 NG ($\sim 8\%$) systems with 2\,344 galaxies. The distributions of virial radius, LOS velocity dispersion, and redshift values for G and NG clusters are shown in Fig.~\ref{fig:distribution_r200_sigma_z}.

We create two ensemble clusters following the procedure outlined in Section~\ref{subsec:Ensemble_Cluster_Methods}, one containing all 588 G systems and the other containing the remaining 54 NG systems. Each ensemble cluster is characterized by a virial radius $\langle r_{200} \rangle$, a velocity dispersion $\langle \sigma_v \rangle$ and a redshift $\langle z \rangle$, which are obtained by taking the average of the $r_{200}$, $\sigma_v$ and $z$, respectively, of the individual clusters that compose the ensemble cluster. The values of $\langle r_{200} \rangle$, $\langle \sigma_v \rangle$ and $\langle z \rangle$ are shown in Table~\ref{tab:parameters_ensemble_G_NG}. Due to the issues mentioned in Section~\ref{subsec:MAMPOSSt_Methods}, galaxies with $R/r_{200} < 0.1$ and $R/r_{200} > 1$ were removed from the sample, resulting in the loss of 3\,326 G galaxies and 310 NG galaxies. The final sample therefore includes 18\,307 G galaxies and 2\,034 NG galaxies. 

\begin{table}
    \centering
    \caption{Average values of the virial radius, $\langle r_{200} \rangle$, velocity dispersion, $\langle \sigma_v \rangle$, and redshift, $\langle z \rangle$, for each ensemble cluster (EC).}
    \label{tab:parameters_ensemble_G_NG}
    \begin{tabular}{cccc} 
        \hline
          EC & $\langle r_{200} \rangle$\,(kpc) & $\langle \sigma_v \rangle$\,(km\,s$^{-1}$) & $\langle z \rangle$\\
        \hline
        G & 1225.39 & 499.14 & 0.0676\\
        NG & 1323.45 & 540.06 & 0.0618\\
        \hline
    \end{tabular}
\end{table}

\subsection{Projected numerical density profiles}
\label{subsec:Recentralization}

\begin{figure*}
        \centering
        \includegraphics[width = \linewidth]{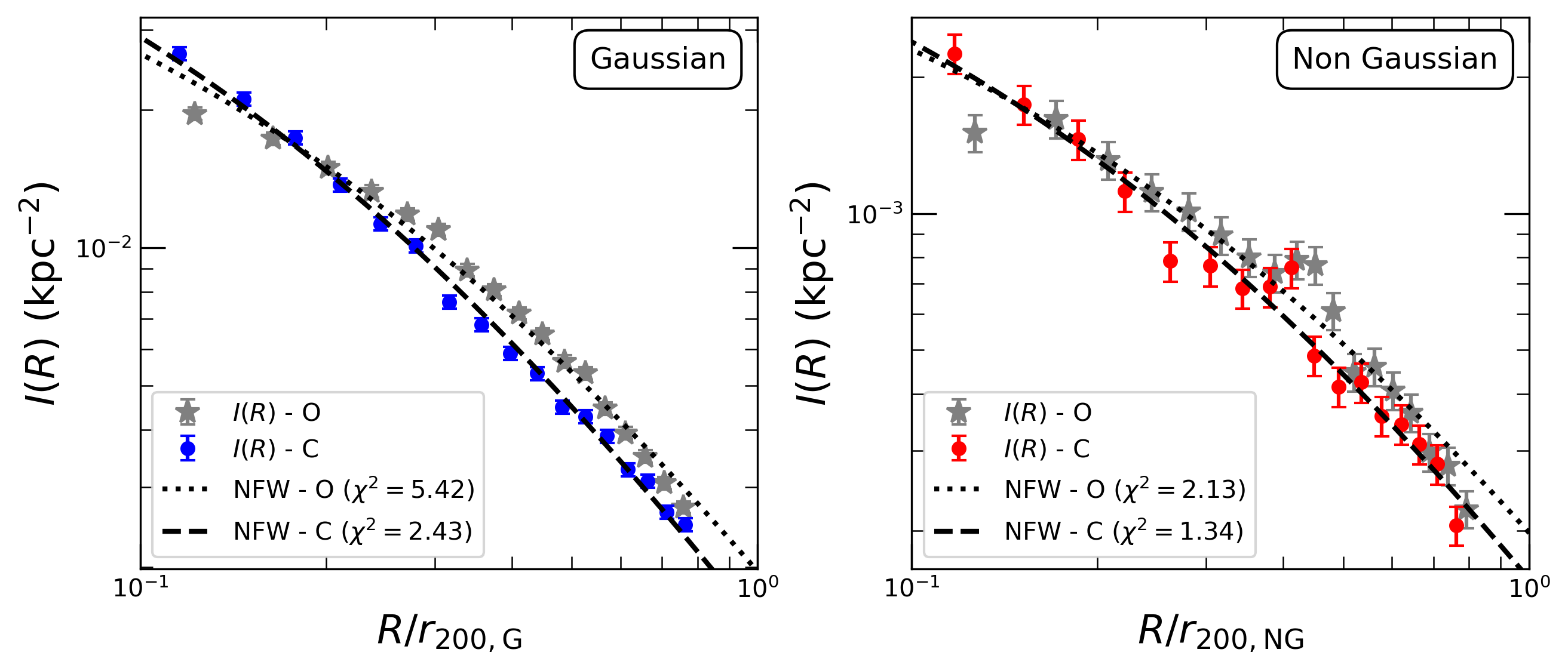}
        \caption{$I(R)$ profile for the G (blue) and NG (red) ensemble clusters. The NFW profile fitted to each profile is represented by the dashed black line. The $I(R)$ profile obtained using the O centre is shown in grey, along with its respective fit represented by the dotted black line (see text for details). The horizontal values in each panel correspond to the central value of each bin.}
        \label{fig:I_fit_G_NG}
    \end{figure*}

    The $I(R)$ profiles of the G (blue) and NG (red) ensemble clusters, estimated considering bins with 5\% of the respective sample size (except for the last bin), are shown in Fig.~\ref{fig:I_fit_G_NG}. We also show in grey the $I(R)$ profiles obtained using the O centre for the position of each individual cluster during the process of creating the ensemble clusters (see Section~\ref{subsec:Identifying_the_clusters_central_positions} for details). We restrict ourselves to showing the profile only until $R/r_{200} = 0.8$, the reason is that not all the clusters contributed with galaxies to all distances in the ensemble cluster. For $R/r_{200} = 0.8$, approximately $95\%$ of the clusters contribute with galaxies; however, this percentage decreases rapidly beyond this radius. As a consequence, the $I(R)$ profile of the ensemble cluster will present an artificial sharp drop in the external region ($R/r_{200} > 0.8$).
    
    We fit these profiles using a NFW profile (see Section~\ref{subsec:Observed_projected_profiles_Methods}). The fit was performed considering only  bins at $R/r_{200} \leq 0.8$. The quality of each fit was quantified by computing $\chi^2$ through the equation
    \begin{equation}
        \chi^2 = \frac{1}{N-2}\sum_i\frac{(I_{o,i}-I_{f,i})^2}{\sigma_{I,i}^2},
    \end{equation}

    \noindent where $I_{o,i}$ is the $i$-th value of the observed $I(R)$ profile, $I_{f,i}$ is the value of the fitted profile at the radial distance $R$, $\sigma_I$ is the uncertainty in the $I_{o,i}$ value, and $N$ is the number of bins in the $I(R)$ profile. The NFW profiles fitted to the $I(R)$ profiles for each ensemble cluster are shown in Fig.~\ref{fig:I_fit_G_NG} for the C (dashed black line) and O (dotted black line) centres. Table~\ref{tab:table_I_fit_G_NG} presents the best-fitting values of $\nu_0$ and $r_\nu$, with their uncertainties, and the $\chi^2$ obtained for each one of the $I(R)$ profiles. As mentioned in Section~\ref{subsec:Identifying_the_clusters_central_positions}, Fig.~\ref{fig:I_fit_G_NG} shows that the $I(R)$ profile depends on the centre choice. The O centre produces $I(R)$ profiles that are flatter in the central region and exhibit a slower decline in the outer region compared to those obtained with the C centre. Additionally, the NFW profile better fits the $I(R)$ profiles obtained considering the C centre, as confirmed by the $\chi^2$ values in Table~\ref{tab:table_I_fit_G_NG}. 

    The scale radius of the NFW profile gives us information about the typical distance of galaxies to the centre of the cluster. Analysing the best-fitting values given in Table~\ref{tab:table_I_fit_G_NG}, we see that galaxies belonging to G clusters tend to present cluster-centric distances slightly lower than those in NG clusters. This can be an indication that G systems are closer to equilibrium, while NG systems are still accreting matter from their surroundings. The $I(R)$ profile of the NG ensemble cluster presents a gap in $R/r_{200} \sim 0.4$. This strange behaviour, not seen in the $I(R)$ profile for the G ensemble cluster, possibly arises due to the presence of substructures in NG systems, which disturb the distribution of galaxies in these clusters.

    \begin{table}
    	\centering
    	\caption{Best-fitting values of $r_\nu$ and $\nu_0$, with their respective uncertainties, and the $\chi^2$ of the fit, for the $I(R)$ profile. The results are shown for the G and NG ensemble clusters considering both definitions for the centre of the individual clusters, namely: Observed (O) and Corrected (C).}
    	\label{tab:table_I_fit_G_NG}
    	\begin{tabular}{ccccccc} 
    		\hline
    		  EC & Centre & $r_\nu$ (kpc) & $\nu_0$ (kpc$^{-3}$) & $\chi^2$\\
    		\hline
    		G & O & $460 \pm 39$ & $2434 \pm 169$ & $5.42$ \\
    		G & C & $304 \pm 17$ & $1594 \pm 63$\phantom{1} & $2.43$ \\
    		NG & O & $594 \pm 95$ & $311 \pm 44$ & $2.13$ \\
    		NG & C & $433 \pm 57$ & $214 \pm 22$ & $1.34$ \\
    		\hline
    	\end{tabular}
    \end{table}

    
\subsection{MAMPOSSt and IJE results for relaxed and non-relaxed clusters}
\label{subsec:Mass_profile_G_NG}
\label{sec:Velocity_anisotropy_profiles_G_NG}

    We execute MAMPOSSt in the \textit{Split} mode for the G and NG ensemble clusters, providing the $r_\nu$ values in Table~\ref{tab:table_I_fit_G_NG}. The complete sample of galaxies in each ensemble cluster was considered as being part of a single tracer population. We provide MAMPOSSt with the redshift of each ensemble cluster (see Table~\ref{tab:parameters_ensemble_G_NG}). Table~\ref{tab:parameters_MAMPOSSt_G_NG} presents the results obtained by MAMPOSSt for the parameters of the $\nu(r)$, $M(r)$ and $\beta(r)$ profiles. The central value of each parameter is given by the 50th quantile of its marginal distribution, while uncertainties are represented by the 5th and 95th percentiles.

\begin{figure}
        \centering
        \includegraphics[width=\linewidth]{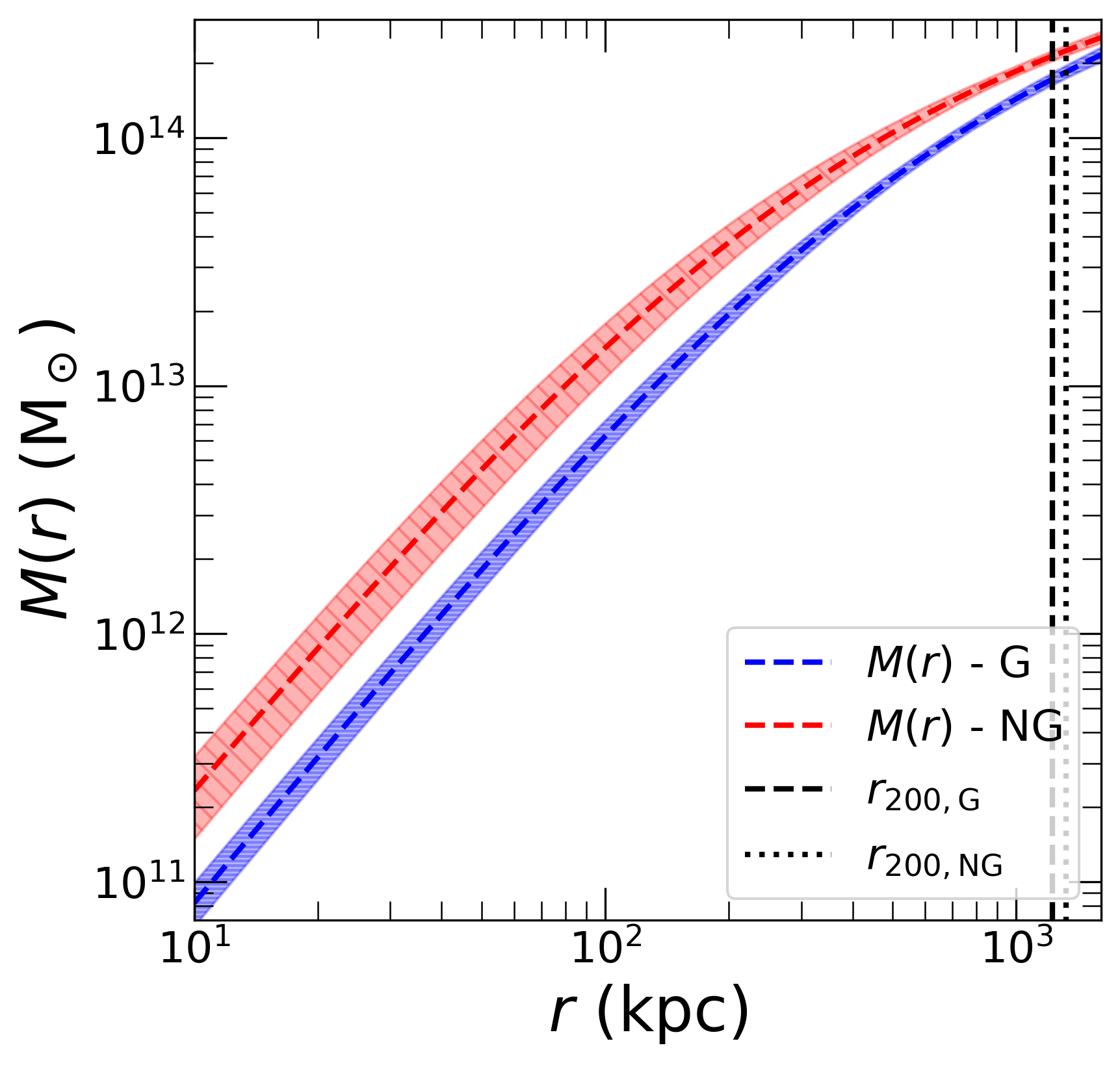}
        \caption{$M(r)$ profile for the G (blue) and NG (red) ensemble clusters obtained from MAMPOSSt. The shaded regions represent the uncertainties in each profile (see text for details). The mean $r_{200}$ values of the G and NG ensemble clusters from Table~\ref{tab:parameters_ensemble_G_NG} are represented, respectively, by the dashed and dotted black lines.}
        \label{fig:Mass_Profiles_G_NG}
    \end{figure}

    The MAMPOSSt solution for the $M(r)$ profile provides \mbox{$r_{200,\rm{G}} = \uncertainties{1122}{50}{26}$}\,kpc and $r_{-2,\rm{G}} = \uncertainties{389}{126}{112}$\,kpc for the G ensemble cluster and $r_{200,\rm{NG}} = \uncertainties{1230}{55}{29}$\,kpc and $r_{-2,\rm{NG}} = \uncertainties{195}{72}{160}$\,kpc for the NG ensemble cluster. The $r_{200}$ values estimated by MAMPOSSt for both ensemble clusters are slightly lower ($\sim9\%$ for G and $\sim8\%$ for NG) than the average values given in Table~\ref{tab:parameters_ensemble_G_NG}. The mass profiles obtained by MAMPOSSt for the G and NG ensemble clusters are shown in Fig.~\ref{fig:Mass_Profiles_G_NG}, respectively, by the blue and red dashed lines. The shaded regions represent the uncertainties in these profiles, estimated from bootstraps and calculated using the normalized interquartile range (NIQR), where $\rm{NIQR}=\rm{IQR}/1.349$. The $r_{200}$ values of the G and NG ensemble clusters from Table~\ref{tab:parameters_ensemble_G_NG} are shown, respectively, by the dashed and dotted black lines. The mass profiles of the G and NG ensemble clusters differs significantly. The $M(r)$ profile of the G ensemble cluster presents a concentration factor of $c_{\rm{G}} = \uncertainties{2.88}{0.68}{1.18}$, while for the NG ensemble cluster we obtain $c_{\rm{NG}} = \uncertainties{6.31}{2.83}{3.58}$. Therefore, the MAMPOSSt solutions suggest that G systems tend to be much less concentrated than NG systems. We also found that NG clusters are slightly more massive than G clusters ($M_{200, \rm{NG}} = 10^{14.33} \,M_\odot$ and $M_{200, \rm{G}} = 10^{14.20} \, M_\odot$), in agreement with the results obtained by \citet{Ribeiro_2011_MNRAS}.

    \begin{table*}
        \centering
        \caption{Values for the parameters obtained from MAMPOSSt for the G and NG ensemble clusters. The value of each parameter is given by the 50th quantile of its marginal distribution, while uncertainties are represented by the 5th and 95th percentiles.}
        \begin{tabular}{ccccccc}
            \hline
             EC & $\log \, r_{200}$ (kpc) & $\log \, r_{-2}$ (kpc) & $\log \, r_{\nu}$ (kpc) & $\beta_0$ & $\beta_\infty$ & $\log \, r_{\beta}$ (kpc) \\
            \hline
                G & $\uncertainties{3.05}{0.02}{0.01}$ & $\uncertainties{2.59}{0.17}{0.11}$ & $\uncertainties{2.48}{0.04}{0.04}$ & $\uncertainties{-0.08}{1.07}{0.39}$ & $\phantom{-}\uncertainties{0.43}{0.26}{0.49}$ & $\uncertainties{2.71}{1.30}{0.73}$ \\
        \addlinespace
                NG & $\uncertainties{3.09}{0.02}{0.01}$ & $\uncertainties{2.29}{0.20}{0.26}$ & $\uncertainties{2.63}{0.10}{0.10}$ & $\uncertainties{-0.61}{1.14}{1.29}$ & $\uncertainties{-0.73}{1.01}{1.16}$ & $\uncertainties{2.42}{1.66}{1.80}$ \\
        \hline
        \end{tabular}
        \label{tab:parameters_MAMPOSSt_G_NG}
    \end{table*} 

\begin{figure}
        \centering
        \includegraphics[width=\linewidth]{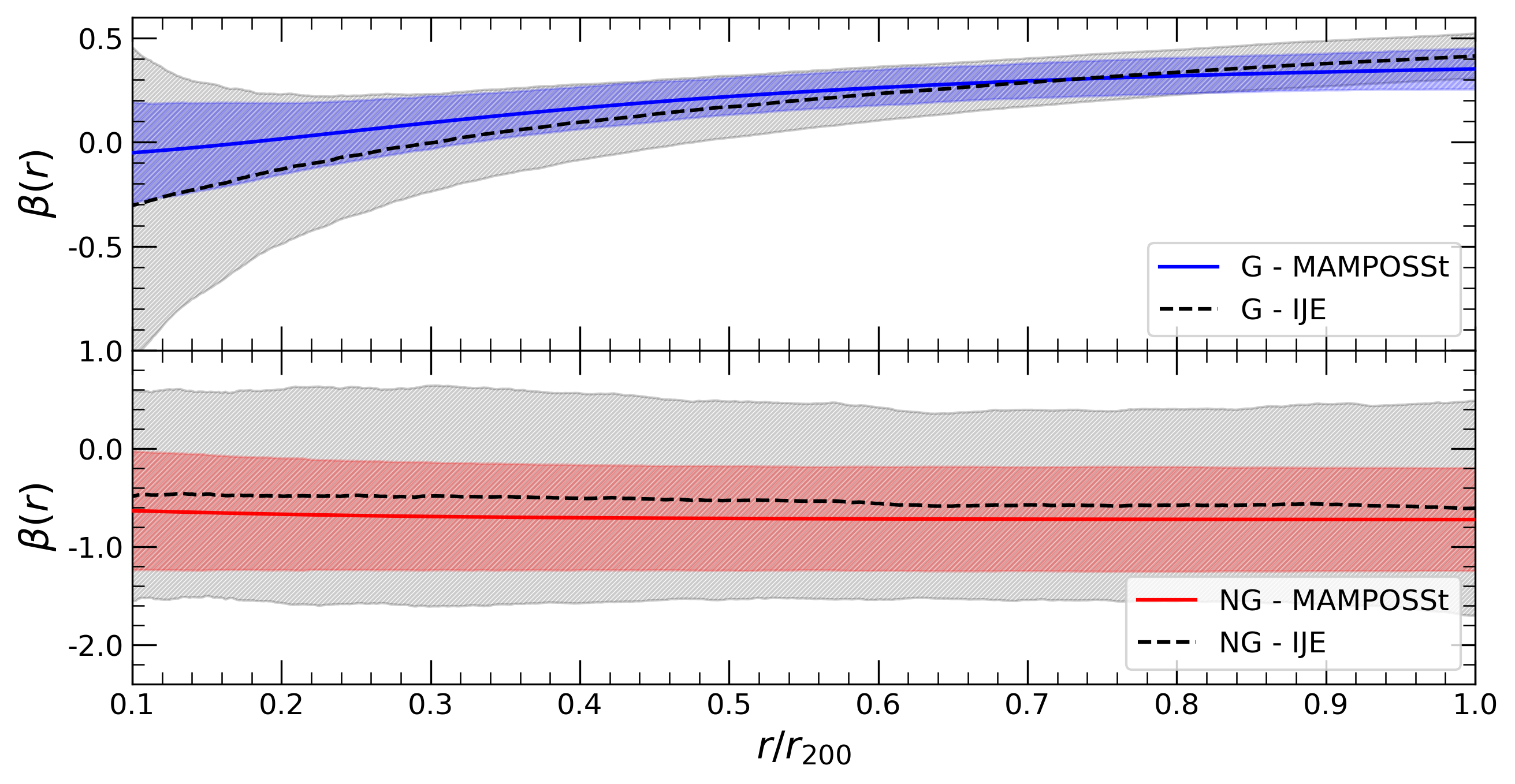}
        \caption{$\beta(r)$ profiles estimated from MAMPOSSt (coloured line) and IJE (dashed black line) for the G (upper panel) and NG (lower panel) ensemble clusters. The uncertainties in the MAMPOSSt and IJE $\beta(r)$ profiles are represented by the coloured and grey shaded regions, respectively. The horizontal axis in each panel provides the radial distance normalized by the respective $r_{200}$ mean value given in Table~\ref{tab:parameters_ensemble_G_NG}.}
        \label{fig:Beta_profiles_G_NG}
    \end{figure}

    The $\beta(r)$ profiles obtained from MAMPOSSt for the G and NG ensemble clusters are shown in Fig.~\ref{fig:Beta_profiles_G_NG}, respectively, in the upper (blue continuum line) and lower (red continuum line) panels. The coloured shaded regions represent the uncertainties in these profiles, estimated from bootstraps and calculated using the NIQR. The average $\beta(r)$ profile estimated by the IJE is given by the dashed black line in each panel and its uncertainties, calculated using the NIQR, are given by the grey shaded region.

    Galaxies in G clusters tend to be characterized by more isotropic orbits in the central regions that become more radial with increasing cluster-centric distances. Considering the uncertainties, the $\beta(r)$ profiles obtained from MAMPOSSt and IJE for the G system seem to agree very well. This behaviour for the $\beta(r)$ profile, with the entire galaxy population being characterized by more isotropic orbits near the centre that become radial outside, agrees with the results obtained by several authors in the literature \citep[e.g.][]{Biviano_2013_A&A, Aguerri_2017_MNRAS, Biviano_2021_A&A}. On the other hand, the shape of the $\beta(r)$ profile for the NG system is completely different. The MAMPOSSt and IJE solutions for the NG ensemble cluster indicate that galaxies in NG systems present tangential orbits at all radial distances. We interpret this unusual behaviour for the $\beta(r)$ profile as an indication that these systems are out of equilibrium and not as a real circular anisotropy in the orbits of galaxies in NG systems.

    The results obtained in this section for both the G and NG ensemble clusters highlight the differences between these two types of systems. Galaxies in G systems are closer to the centre regions and present a more regular spatial distribution (no unusual features in the $I(R)$ profile), when compared to galaxies in NG systems. The $M(r)$ and $\beta(r)$ profiles estimated for these systems are also very different, indicating that G and NG systems are characterized by different dynamics. Therefore, if we understand Gaussian systems as objects closer to the equilibrium condition \citep[e.g.][]{Ribeiro_2011_MNRAS, Ribeiro_2013_MNRAS}, this implies that NG systems are still far from equilibrium. In addition, a specific analysis to investigate what causes the observed behaviour in the $\beta(r)$ profile of NG systems is necessary but is beyond the scope of this work and will be addressed in a future work.
    
    Finally, the significant differences between the velocity anisotropy profiles are of particular importance for this work. These differences underscore the importance of removing systems far from equilibrium in order to conduct a reliable orbital analysis. However, due to the complexity of the signatures of non-equilibrium in clusters \citep[see e.g.][]{Einasto_2012_A&A}, it is important to use multiple tools to identify them. 


\subsection{Removing non-relaxed clusters from the sample}
\label{subsec:Removing_nonrelaxed_clusters_from_the_sample}

The gaussianity of the LOS velocity distribution of a cluster is an indicator of its dynamical state, but it is not the only one \citep[e.g.][]{Einasto_2012_A&A}. Clusters close to equilibrium are expected to exhibit a symmetric and spherical distribution of galaxies \citep[e.g.][]{Dubinski_1994_ApJ,Kazantzidis_2004_ApJL}. Therefore, the asymmetry of the distribution of galaxies or their degree of flattening can be used as indicators of non-equilibrium for galaxy clusters. Additionally, systems that undergo a merging process also violate the equilibrium condition. We verified that the removal of NG clusters alone was insufficient to eliminate all highly asymmetric, flattened, and merging clusters from our sample of G clusters. Thus, beyond removing NG systems to ensure a reliable orbital analysis, we also conducted an additional analysis to identify and exclude highly asymmetric, flattened, and merging clusters. In the following, we discuss the methods and criteria used.

We adopt the $\beta$ test, presented by \citet{West_1988_ApJ}, to identify and remove highly asymmetric clusters from our sample. Let $d_i$ be the average projected distance between the $i$ galaxy and its five nearest neighbours in projection and $d_o$ the average projected distance between a point diametrically opposite to the position of the galaxy, with respect to the centre of the cluster, and its five nearest neighbours. Then, the $\beta$ parameter is defined as
\begin{equation}
\beta = \log_{10}(d_0/d_i).
\end{equation}

\noindent The asymmetry of a cluster is quantified by the average value of $\beta$, $\langle \beta \rangle$, estimated over all galaxies. For symmetric clusters, $\langle \beta \rangle \sim 0$ is expected, while asymmetric clusters tend to present $\langle \beta \rangle > 0$. 

We apply the $\beta$ test for the 588 G clusters in our sample. To identify and exclude highly asymmetric clusters from the sample we apply a bootstrap procedure with 10\,000 resamplings for each cluster, estimating the $\langle \beta \rangle$ value in each draw. The procedure consists of drawing random positions for the galaxies, keeping fixed their distances to the centre, with the aim of breaking possible asymmetries in the cluster. A cluster is classified as asymmetric if the fraction of all resamplings where $\langle \beta \rangle$ is larger than the measured one is lower than $5\%$. Applying this criterion, we identify 61 asymmetric clusters in our sample, containing a total of 3\,375 galaxies.

The identification and removal of highly flattened clusters from our sample were performed using a criterion based on their apparent ellipticities. To estimate the ellipticity of a cluster, we apply the method of moments presented by \citet{Rhee_1991_A&AS}. In this method, the moments of the galaxy distribution are defined as 
\begin{equation}
    \mu_{ab} = \dfrac{\sum(x-x_0)^a(y-y_0)^b}{N_{\rm{gal}}},
\end{equation}

\noindent where $x$ and $y$ are the galaxy coordinates, $(x_0,y_0)$ represent the coordinates of the cluster centre, and $N_{\rm{gal}}$ is the number of galaxies in the cluster. Then, the cluster elongation, $e$, defined as
\begin{equation}
    e = \dfrac{\sqrt{(\mu_{20}-\mu_{02})^2+4\mu_{11}^{2}}}{\mu_{20}+\mu_{02}},
\end{equation}

\noindent can be used to estimate the standard ellipticity, $\epsilon$, through the equation 
\begin{equation}
    \epsilon = 1- \sqrt{\dfrac{1-e}{1+e}}.
\end{equation}

\noindent The uncertainties on $\epsilon$ are estimated by drawing randomly galaxies from the cluster (keeping the size, but allowing repetitions) and estimating the $\epsilon$ value for each draw. This process is executed 10\,000 times for each cluster, and the standard deviation ($\sigma_\epsilon$) and the 5th and 95th percentiles of the distribution of values of $\epsilon$ are computed at the end. The threshold value to remove flattened clusters was obtained by simulating $100$ clusters, like in the $\beta$ test, for each cluster in our sample and estimating the $\epsilon$ and $\sigma_\epsilon$ values for each simulated cluster. The mean value of $\epsilon$ for the observed and simulated clusters is, respectively, $\langle \epsilon_{o} \rangle = \uncertainties{0.41}{0.31}{0.23}$ and $\langle \epsilon_{s} \rangle = \uncertainties{0.17}{0.12}{0.20}$. The $\langle \epsilon \rangle$ value is estimated using $\langle \epsilon \rangle = \sum_i (\epsilon_i w_i)/ \sum_i w_i$, with $w_i = 1/\sigma_{\epsilon,i}^2$, and the uncertainties correspond to the 5th and 95th percentiles of the respective distributions. We find that only $\sim 5\%$ of the simulated clusters have $\epsilon>0.37$. Therefore, we adopt this value as the threshold value to classify a cluster as flattened or not. This criterion implies the identification of 98 flattened clusters with a total of 3\,080 galaxies.

Finally, clusters in the process of merging also violate the equilibrium condition even if such clusters are not detected as unrelaxed systems by the methods previously described. Therefore, we remove possible merging systems from our sample of G clusters. It is important to mention that the aim of this step is not to use a robust method to identify and remove, with high accuracy, all the merging systems from our sample, but to reduce the contribution of such systems from our sample of G clusters. To achieve this, we use the merger catalogue presented by T17. As mentioned in Section~\ref{sec:Data}, T17 identified 498 potentially merging systems in their group catalogue. We verify that, among our 588 G clusters, 140 are part of merging systems. These 140 systems contain a total of 6\,500 galaxies.

We remove from our sample of 588 G clusters all the systems identified as asymmetric, flattened or in process of merging. This results in a final sample of 336 G clusters containing 10\,898 galaxies. Some clusters are removed from the G sample by more than one method simultaneously; therefore, the number of clusters removed is not equal to the total numbers of asymmetric, flattened and merging systems.
    

\section{Galaxy populations}
\label{sec:Galaxy_populations}


    In this section, we split galaxies belonging to G clusters into galaxy populations and analyse their kinematical and dynamical properties. 


\subsection{Classification into populations}
\label{subsec:Classification_Pops}

    The 10\,898 galaxies belonging to the 336 G clusters were classified into populations according to their dominant gas ionization mechanism, using the criteria discussed in Section~\ref{subsec:Identification_of_the_gas_ionization_source}. The number of galaxies classified in each population is 6\,923 Q, 2\,425 SF, 755 T, and 218 AGN galaxies, respectively. We lost a total of 577 galaxies during the classification process, including 77 galaxies with an emission line pattern that does not meet our classification criteria, 405 galaxies for which emission line measurements have not been performed, and 95 galaxies without SDSS spectroscopic observations altogether.

    These 336 G clusters were subsequently used to create an ensemble cluster, following the procedure outlined in Section~\ref{subsec:Ensemble_Cluster_Methods}. Only galaxies located at cluster-centric distances in the range $0.1 \leq R/r_{200} \leq 1.0$ are included in the ensemble cluster (see Section~\ref{subsec:MAMPOSSt_Methods}); this results in the loss of 1\,429 galaxies. The remaining 8\,892 galaxies in the ensemble cluster are distributed as follows: 5\,890 Q, 2\,142 SF, 672 T, and 188 AGN galaxies. The virial radius, velocity dispersion and redshift of the ensemble cluster (see Section~\ref{subsec:Ensemble_Cluster_Methods}) are, respectively, $\langle r_{200} \rangle = 1194.73$\,kpc, $\langle \sigma_v \rangle = 485.74$\, km\,s$^{-1}$ and $\langle z \rangle = 0.071824$.


\subsection{Projected numerical density profiles}
\label{subsec:I_profiles_Pops}
 
\begin{figure*}
        \centering
        \includegraphics[width = \linewidth]{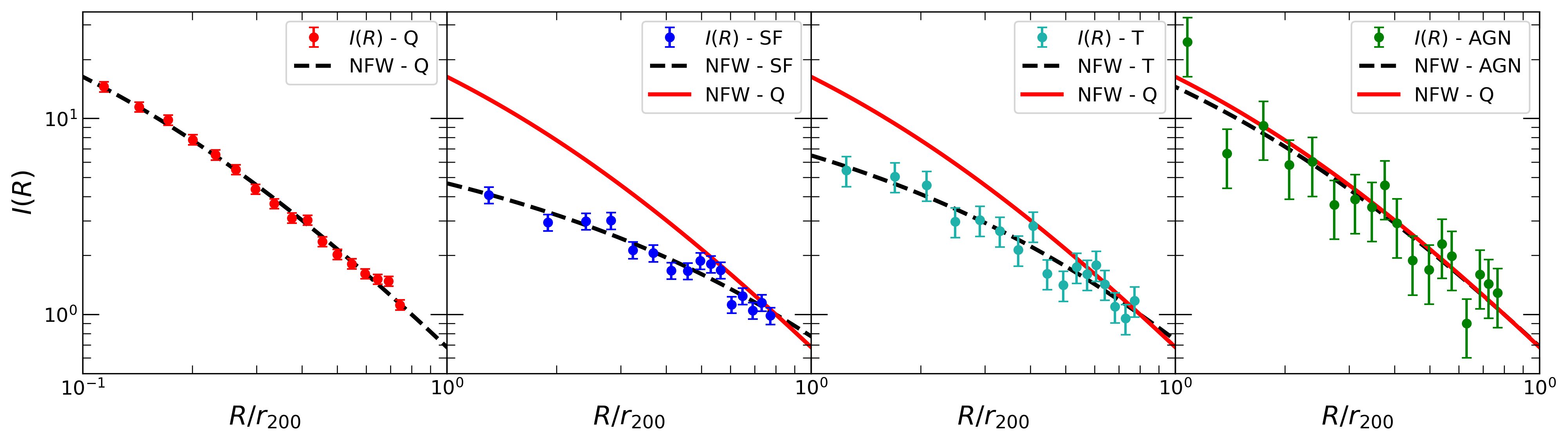}
        \caption{The $I(R)$ profile (coloured dots) and the fitted NFW profile (dashed black line), both normalized by the value of the fitted NFW profile in $R/r_{200} = 0.8$, for each galaxy population, namely: Q (red), SF (blue), T (cyan), and AGN (green). The fitted NFW profile of the Q population is shown in the panels of the remaining populations to better compare the slope of the profiles (red solid line). The horizontal axis in each panel provides the radial distance normalized by the $\langle r_{200} \rangle$ value given in Section~\ref{subsec:Classification_Pops}, and the horizontal values in each panel correspond to the central value of each bin.}
        \label{fig:I_profile_Pops}
    \end{figure*}

    The $I(R)$ profiles, with their respective uncertainties, estimated for each population are shown in Fig.~\ref{fig:I_profile_Pops} (coloured dots). These profiles have been estimated in bins containing $5 \%$ of the respective population sample size. The fitted NFW profile is represented by the dashed black line in each panel of Fig.~\ref{fig:I_profile_Pops}. The best-fitting values for $r_\nu$ and $\nu_0$, with their uncertainties, obtained for each population are given in Table~\ref{tab:table_I_fit_Pops}, along with the $\chi^2$ of the fit. The $I(R)$ and NFW profiles are normalized by the value of the fitted NFW profile at $R/r_{200} = 0.8$. In order to better compare the slope of the profiles, the NFW profile of the Q population is included in the panels of the other populations (red solid line).

    Analysing the $I(R)$ profiles, we observe that the Q and AGN populations exhibit very similar profiles. Galaxies belonging to these two populations tend to be located closer to the central regions when compared to SF and T galaxies. The SF and T populations also present similar $I(R)$ profiles. However, in this case it is possible to identify a different spatial distribution, with SF galaxies presenting the largest typical cluster-centric distances. Transition galaxies tend to inhabit the intermediate distances between those galaxies closest to the centre (Q and AGN) and the more distant SF galaxies.
        
    \begin{table}
    	\centering
    	\caption{Best-fitting values of $r_\nu$ and $\nu_0$, with their respective uncertainties, and the $\chi^2$ of the fit, for the $I(R)$ profiles of the galaxy populations.}
    	\label{tab:table_I_fit_Pops}
    	\begin{tabular}{ccccccc} 
    		\hline
    		  Pop. & $r_\nu$ (kpc) & $\nu_0$ (kpc$^{-3}$) & $\chi^2$\\
    		\hline
    		Q & $\phantom{1}229 \pm 16\phantom{1}$ & $445 \pm 19\phantom{1}$ & $0.95$ \\
    		SF & $ 1211 \pm 215$ & $621 \pm 127$ & $1.47$ \\
    		T & $\phantom{1}752 \pm 142$ & $124 \pm 23 \phantom{1}$ & $0.79$ \\
    		AGN & $ \phantom{1} 267 \pm 86 \phantom{1}$ & $\phantom{1}16 \pm 3\phantom{11}$& $0.88$ \\
    		\hline
    	\end{tabular}
    \end{table}


\subsection{Mass and anisotropy profiles}
\label{subsec:Mass_profile_Pops}
\label{subsec:Beta_profiles_Pops}

    We execute MAMPOSSt for the ensemble cluster considering each galaxy population as an independent tracer of the cluster potential. The NFW profile was considered for both $\nu(r)$ and $M(r)$ profiles and the generalized Osipkov-Merritt profile was used for the $\beta(r)$ profile. The execution was performed using both the \textit{Split} ($r_\nu$ fitted externally) and \textit{TAND} ($r_\nu = r_\beta$) modes of MAMPOSSt (see Section~\ref{subsec:MAMPOSSt_Methods}). We provided for MAMPOSSt the $r_\nu$ values from Table~\ref{tab:table_I_fit_Pops}, and the redshift of the ensemble cluster given in Section~\ref{subsec:Classification_Pops}. The MAMPOSSt results obtained for the galaxy populations are shown in Table~\ref{tab:Results_MAMPOSSt_Pops}. The central value of each parameter is given by the 50th quantile of its marginal distribution, while uncertainties are represented by the 5th and 95th percentiles.

    The MAMPOSSt solution for the $M(r)$ profile provides \mbox{$r_{200} = \uncertainties{1047}{24}{25}$\,kpc} and $r_{-2} = \uncertainties{347}{84}{100}$\,kpc. The $r_{200}$ value obtained from MAMPOSSt is slightly lower ($\sim 12\%$) than the real value ($\langle r_{200} \rangle = 1194.73$\,kpc). The concentration factor of the mass profile obtained is $c = \uncertainties{3.02}{0.72}{0.98}$, while the mass contained within the virial radius is estimated to be $M_{200} = 10^{14.13} \,M_\odot$. 
    
    \begin{table}
        \centering
        \caption{MAMPOSSt results for the $\nu(r)$ and $\beta(r)$ profiles of the galaxy populations. The central value of each parameter is given by the 50th quantile of its marginal distribution, while uncertainties are represented by the 5th and 95th percentiles. We assumed $r_\nu = r_\beta$ in the MAMPOSSt execution.}
        \begin{tabular}{cccc}
            \hline
             Pop. & $\log \, r_{\nu}$ (kpc) & $\beta_0$ & $\beta_\infty$ \\
            \hline
            Q & $\uncertainties{2.35}{0.05}{0.05}$ & $\phantom{-}\uncertainties{0.01}{0.87}{0.65}$ & $\uncertainties{0.26}{0.25}{0.20}$ \\
        \addlinespace
            SF & $\uncertainties{3.00}{0.11}{0.11}$ & $\uncertainties{-0.05}{0.39}{0.29}$ & $\uncertainties{0.66}{0.17}{0.14}$ \\
        \addlinespace
            T & $\uncertainties{2.86}{0.13}{0.13}$ & $\uncertainties{-0.52}{0.97}{0.63}$ & $\uncertainties{0.70}{0.26}{0.20}$ \\
        \addlinespace
            AGN & $\uncertainties{2.56}{0.23}{0.24}$ & $\uncertainties{-1.45}{3.72}{1.55}$ & $\uncertainties{0.84}{0.33}{0.14}$ \\
        \hline
        \end{tabular}
        \label{tab:Results_MAMPOSSt_Pops}
    \end{table} 



\begin{figure*}
        \centering
        \includegraphics[width = \linewidth]{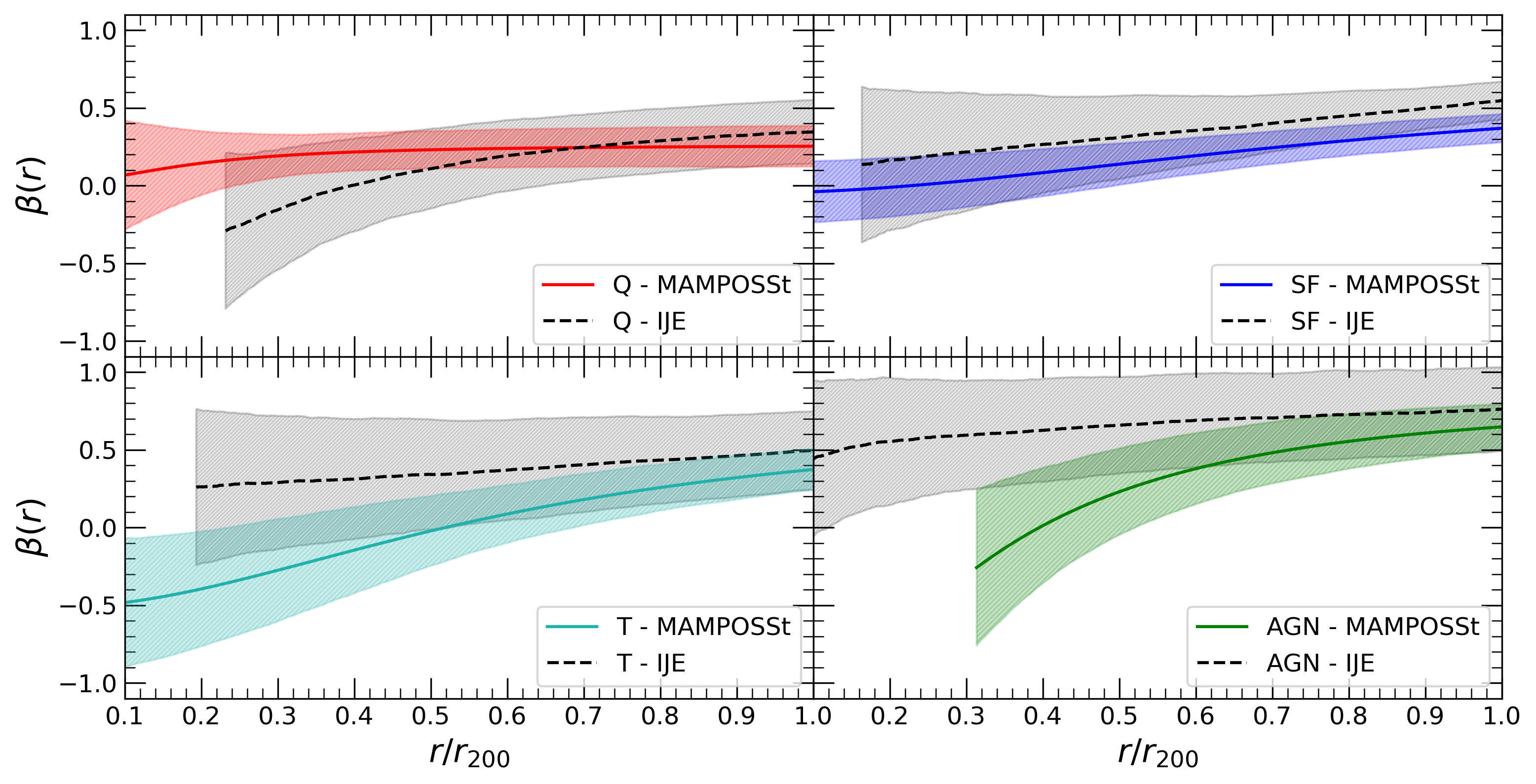}
        \caption{
        The $\beta(r)$ profiles estimated from MAMPOSSt (coloured solid line) and IJE (dashed black line) for the Q (upper left panel), SF (upper right panel), T (bottom left panel) and AGN (bottom right panel) galaxy populations. The uncertainties in the MAMPOSSt and IJE $\beta(r)$ profiles are given, respectively, by the coloured and black shaded regions (see text for details). The $\beta(r)$ profiles are shown only in the regions where their uncertainties are lower than $0.5$. The horizontal axis in each panel provides the radial distance normalized by the $\langle r_{200} \rangle$ value given in Section~\ref{subsec:Classification_Pops}.}
        \label{fig:Beta_profile_Pops}
    \end{figure*}

    The MAMPOSSt and IJE $\beta(r)$ profiles obtained for the galaxy populations are shown in Fig.~\ref{fig:Beta_profile_Pops}. The MAMPOSSt $\beta(r)$ profiles of the Q, SF, AGN and T populations are given, respectively, by the red, blue, green, and cyan solid lines. The uncertainties in the MAMPOSSt $\beta(r)$ profiles, given by the respective coloured shaded regions, are estimated from bootstraps (see Section~\ref{subsec:Mass_profile_G_NG} for details). The $\beta(r)$ profile estimated from IJE for each population is represented by the dashed black line in each panel of Fig.~\ref{fig:Beta_profile_Pops}. The uncertainties in the IJE $\beta(r)$ profiles, estimated following the procedure described in Section~\ref{sec:Inversion_of_the_Jeans_Equations_Methods}, are given by the black shaded regions. Due to the small number of tracers in each galaxy population, the uncertainties in the estimated $\beta(r)$ profiles can be high. This problem is especially relevant in the internal regions where the available information is poorer and we have less confidence in the results. Therefore, we only show the MAMPOSSt and IJE $\beta(r)$ profiles in the radial regions where the uncertainties in these profiles are lower than $0.5$. This cut-off limit of $0.5$ was arbitrarily chosen to remove poorly constrained regions from Fig.~\ref{fig:Beta_profile_Pops} while still maintaining a sufficiently large spatial coverage of the $\beta(r)$ profiles.

    According to the MAMPOSSt solution, the Q population is characterized by isotropic orbits in the inner regions ($\beta_0 = \uncertainties{0.01}{0.87}{0.65}$) that become increasingly radial with the increase of the cluster-centric distances, reaching $\beta = 0.25$ at $r/r_{200} = 1.0$. The rapid transition from the internal isotropic orbits to the more radial orbits in the outskirts of the clusters presented by the Q population is due to the lower value of $r_\beta$, which is fixed to be equal to $r_\nu$. The IJE $\beta(r)$ profile of the Q population indicates orbits that are slightly more tangential in the inner region ($\beta = -0.29$ at $r/r_{200} = 0.23$) when compared to the MAMPOSSt solution, but both solutions agree at larger radii. We tentatively interpret the differences between the MAMPOSSt and IJE solutions at low radial distances in Section~\ref{subsec:Presence_of_central_galaxies}.

    Galaxies belonging to the SF population are characterized by isotropic orbits in the inner regions ($\beta_0 = \uncertainties{-0.05}{0.39}{0.29}$) that become more radially elongated at larger radii ($\beta = 0.37$ at $r/r_{200} = 1.0$). This trend on the $\beta(r)$ profile is observed for both MAMPOSSt and IJE solutions. However, the IJE solution indicates more radial orbits at all radii when compared to the MAMPOSSt solution, reaching $\beta = 0.55$ at $r/r_{200} = 1.0$. The explanation for this behaviour is related to an excess of kinetic energy that the SF population seems to possess, as discussed in Section~\ref{subsec:SigmaP_profiles_Pops}. Comparing the SF and Q populations, we observe that the galaxies of both populations are characterized by isotropic orbits in the central regions, but the SF galaxies reach marginally more radial orbits in the surroundings of the clusters.

    Transition galaxies, according to the MAMPOSSt solution, present tangential orbits ($\beta_0 = \uncertainties{-0.52}{0.97}{0.63}$) in the inner regions ($r/r_{200} < 0.5$). These orbits become isotropic at $r/r_{200} \sim 0.5$ and more radially elongated at greater radial distances, reaching $\beta = 0.37$ at $r/r_{200} = 1.0$. On the other hand, in the IJE solution, transition galaxies have radial orbits in the central regions ($\beta = 0.26$ at $r/r_{200} = 0.19 $) that become even more radial with the increase in the cluster-centric distances. Taking uncertainties into account, both solutions are consistent with each other and suggest more isotropic orbits in the inner region, becoming more radial in the outer region. This same behaviour is also observed for the AGN population. The MAMPOSSt solution for $\beta(r)$ indicates slightly tangential orbits in the inner regions ($\beta = -0.27$ at $r/r_{200} = 0.31$) and radial orbits at larger radii ($\beta = 0.65$ at $r/r_{200} = 1.0$). Conversely, the IJE $\beta(r)$ profile shows very radial orbits in all radial distances ($\beta = 0.44$ at $r/r_{200} = 0.1$ to $\beta = 0.76$ at $r/r_{200} = 1.0$). However, considering the uncertainties, both solutions are consistent with isotropic orbits in the central region, becoming increasingly radial at larger radii.

    Finally, it is worth noting that, considering the uncertainties, the $\beta(r)$ profiles of all galaxy populations are consistent with isotropic orbits in the inner regions, which become increasingly radial with increasing cluster-centric distances.

\subsection{Projected velocity dispersion profiles}
\label{subsec:SigmaP_profiles_Pops}

\begin{figure*}
        \centering
        \includegraphics[width = \linewidth]{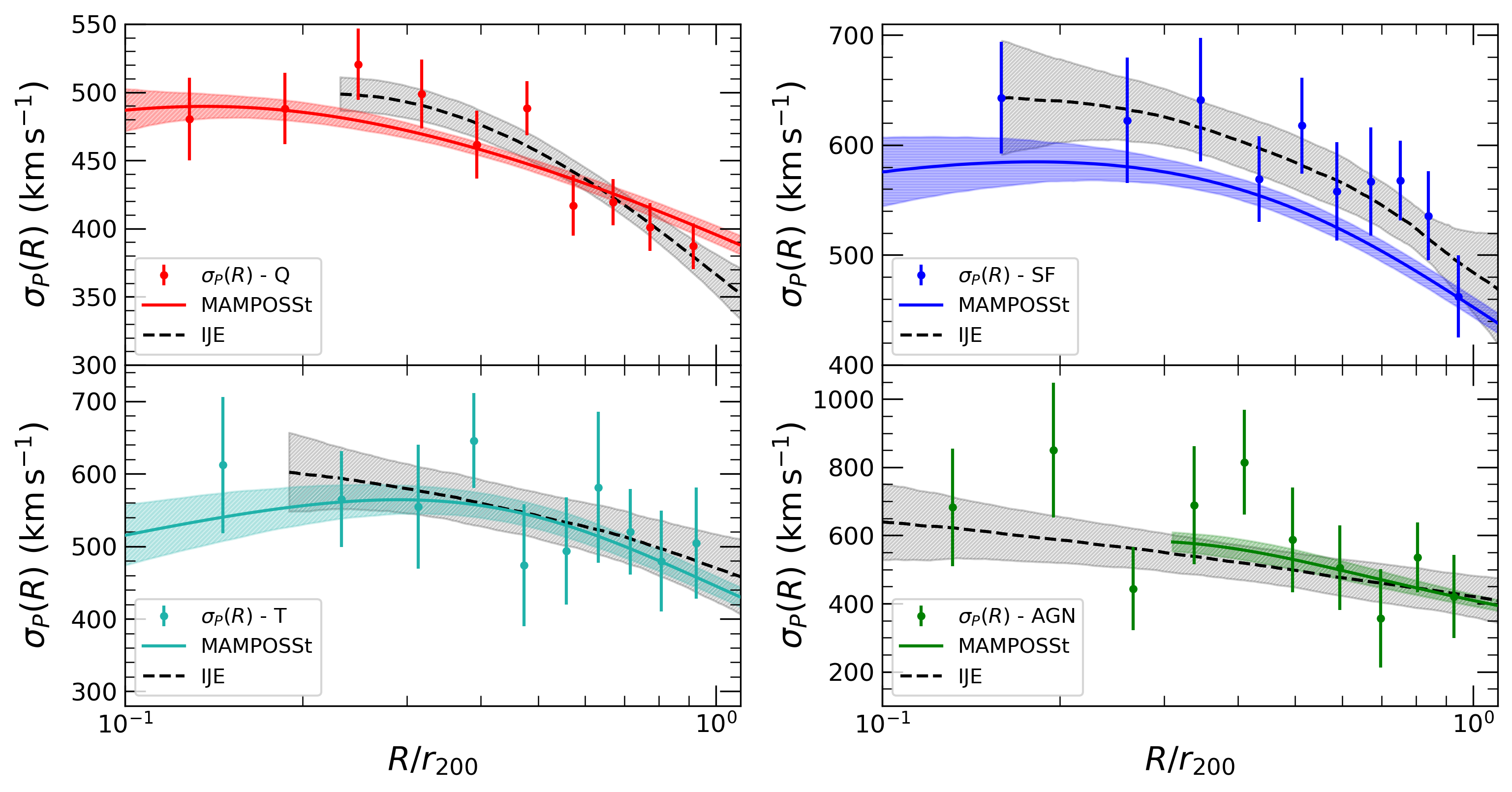}
        \caption{The observed $\sigma_P(R)$ profiles for the Q (red dots, upper left panel), SF (blue dots, upper right panel), T (cyan dots, bottom left panel), and AGN (green dots, bottom right panel) galaxy populations. The MAMPOSSt (coloured solid line) and the IJE (dashed black line) $\sigma_P(R)$ profiles are shown, with their uncertainties represented by the coloured and grey shaded regions, respectively. The horizontal axis in each panel provides the radial distance normalized by the $\langle r_{200} \rangle$ value given in Section~\ref{subsec:Classification_Pops}, and the horizontal values in each panel correspond to the central value of each bin.}
        \label{fig:SigmaP_profile_Pops}
    \end{figure*}

    The observed $\sigma_P(R)$ profiles of the Q, SF, AGN and T galaxy populations are shown in Fig.~\ref{fig:SigmaP_profile_Pops}. The Q population tends to be characterized by lower velocity dispersions when compared to the other populations. There is a clear separation between the $\sigma_P(R)$ profiles of the Q and SF populations, with SF galaxies presenting higher velocity dispersions at all radii. AGN and T galaxies also tend to exhibit higher values of velocity dispersion when compared to quiescent galaxies. It is difficult to compare the $\sigma_P(R)$ profile of the SF population with those of the AGN and T populations due to the high uncertainties in the AGN and T $\sigma_P(R)$ profiles, which are caused by the small number of tracers in these populations. However, the SF population appears to have slightly higher values of velocity dispersion compared to the T population. Unfortunately, the high radial variability of the AGN $\sigma_P(R)$ profile prevents any meaningful comparison with the SF $\sigma_P(R)$ profile.

    To verify the quality of the MAMPOSSt results, we estimate the $\sigma_P(R)$ profile for each galaxy population using the MAMPOSSt solutions for the $M(r)$ and $\beta(r)$ profiles, with its uncertainties estimated from bootstraps. These profiles are given by the coloured solid lines in Fig.~\ref{fig:SigmaP_profile_Pops}, with their uncertainties represented by the respective coloured shaded regions. The median $\sigma_P(R)$ profile from the IJE for each population, estimated from the profiles used to derive the respective $\beta(r)$ profile (see Section~\ref{sec:Inversion_of_the_Jeans_Equations_Methods}), is represented by the dashed black line in Fig.~\ref{fig:SigmaP_profile_Pops}, and its uncertainties by the grey shaded regions. Both MAMPOSSt and IJE $\sigma_P(R)$ profiles are shown only in the regions where the uncertainties in their respective $\beta(r)$ profiles are lower than 0.5 (see Section~\ref{subsec:Beta_profiles_Pops}). To quantify the quality of the solutions from MAMPOSSt and IJE in reproducing the observed $\sigma_P(R)$ profile, we compute the weighted average deviation $SD_W$ between the observed $\sigma_P(R)$ profile and the MAMPOSSt and IJE profiles in Fig.~\ref{fig:SigmaP_profile_Pops}, i.e. 
    \begin{equation}\label{eq:SD_W}
        SD_W = \frac{\sum_i w_i (O_i-I_i)}{\sum_iw_i},
    \end{equation}
    
    \noindent where $O_i$ is the $i$-th observed $\sigma_P(R)$ value, $I_i$ is the value of the interpolated MAMPOSSt/IJE $\sigma_P(R)$ profile at the radial distance $R$, and $w_i = 1/\sigma_i^2$, with $\sigma_i$ being the uncertainty in the $O_i$ value. The uncertainty in $SD_W$ is given by $\sigma_{SD_W} = 1/\sqrt{\sum_i w_i}$. To keep consistency, the $SD_W$ was estimated considering only bins where both MAMPOSSt and IJE $\sigma_P(R)$ profiles are available. The results obtained are shown in Table~\ref{tab:sdw_Pops}.

    The observed $\sigma_P(R)$ profile of the Q population is well reconstructed for both MAMPOSSt and IJE solutions for $\beta(r)$. For this population, the $SD_W$ values do not indicate any statistically significant improvement of one solution over another. This result suggests that the functional form chosen for $\beta(r)$ in the MAMPOSSt execution is sufficient to describe the orbits of the Q population. Additionally, there are no signatures for this population that indicate a higher degree of non-equilibrium, thereby implying that quiescent galaxies are likely close to equilibrium. However, it is interesting to observe that the MAMPOSSt $\sigma_P(R)$ profile seems to ignore an apparent internal curvature ($R/r_{200} \lesssim 0.3$) present in the observed profile. This internal curvature is likely responsible for the more tangential orbits observed in the IJE $\beta(r)$ profile of the Q population shown in Fig.~\ref{fig:Beta_profile_Pops}. 
            
    Contrary to the results for the Q population, MAMPOSSt does not reproduce the observed $\sigma_P(R)$ profile of the SF population. The entire $\sigma_P(R)$ profile obtained from the MAMPOSSt solution appears to be shifted down. On the other hand, the IJE $\sigma_P(R)$ profile reproduces very well the observed profile. This result is confirmed by the $SD_W$ values in Table~\ref{tab:sdw_Pops}, which show that the deviation between the observed and MAMPOSSt $\sigma_P(R)$ profiles is higher than $3\sigma_{SD_W}$. This result explains the origin of the difference between the MAMPOSSt and IJE solutions for the $\beta(r)$ profile of the SF population (Fig.~\ref{fig:Beta_profile_Pops}). The fact that MAMPOSSt was unable to find equilibrium solutions for $M(r)$ and $\beta(r)$ that are able to reproduce the observed $\sigma_P(R)$ profile suggests that the galaxies belonging to the SF population are not in equilibrium within their clusters. There seems to be an excess of kinetic energy in this population that cannot be explained by MAMPOSSt when considering only equilibrium solutions. To account for the higher velocity dispersions observed in the SF population, the IJE method provided as a solution orbits that are more radial at all radii.

    Both MAMPOSSt and IJE $\sigma_P(R)$ profiles reproduce well the observed $\sigma_P(R)$ profile of the T population. According to the $SD_W$ values, both profiles fall within $1\sigma_{SD_W}$ and there is no preference for either of the two solutions. The MAMPOSSt profile presents an inner curvature not visible in the IJE profile. This internal curvature, similar to that of the Q population, is likely responsible for the more tangential orbits observed in the $\beta(r)$ profile. However, it is difficult to conclude which solution better reproduces the observed $\sigma_P(R)$ profile due to the high uncertainties in these inner regions. The $\sigma_P(R)$ profile of the AGN population, like that of the T population, is well reproduced by both MAMPOSSt and IJE profiles. The $SD_W$ values do not favour any of the solutions. The MAMPOSSt solution for the $\sigma_P(R)$ profile presents a curvature\footnote{The curvature in the $\sigma_P(R)$ profile of the MAMPOSSt solution is not visible in Fig.~\ref{fig:SigmaP_profile_Pops} due to the cut imposed by the uncertainties in the $\beta(r)$ profile.} in the inner region ($R/r_{200} < 0.4$), and this produces the more tangential orbits observed in that region. On the other hand, the radial orbits in the inner region observed for the IJE solution occur because its $\sigma_P(R)$ profile does not exhibit such internal curvature. Again, as for the T population, it is difficult to argue in favour of one solution over the other due to the high uncertainties in the $\sigma_P(R)$ profile. 

    Although the MAMPOSSt and IJE solutions for $\beta(r)$ appear to disagree in the inner regions for the AGN and T populations, when considering the uncertainties, both are consistent with isotropic orbits in the inner regions that become more elongated with increasing cluster-centric distances. In addition, we do not observe any measurable displacement between the MAMPOSSt and IJE solutions in reproducing the observed $\sigma_P(R)$ profile, as is the case for the SF population. This can be understood as an indication that galaxies belonging to these two populations are likely closer to equilibrium than SF galaxies.
        
    \begin{table}
        \centering
        \caption{$SD_W$ values for the MAMPOSSt and IJE $\sigma_P(R)$ profiles and the uncertainty in the $SD_W$ value for the Q, SF, AGN, and T populations.}
        \label{tab:sdw_Pops}
        \begin{tabular}{cccccc} 
            \hline
              Pop. & $SD_{W,\mathrm{MAMPOSSt}}$ & $SD_{W,\mathrm{IJE}}$ & $\sigma_{SD_W}$\\
            \hline
            Q &  $\phantom{0}2.9$ & $\phantom{0}2.2$ & $\phantom{0}7.3$ \\
            SF & $46.1$ & $\phantom{0}3.8$ & $14.2$  \\
            T &  $14.5$ & $\phantom{0}0.4$ & $24.4$  \\
            AGN &  $51.4$ & $66.1$ & $50.3$  \\
            \hline
        \end{tabular}
    \end{table}

    \subsection{Quiescent population and the presence of central galaxies}
    \label{subsec:Presence_of_central_galaxies}

    The internal curvature present in the $\sigma_P(R)$ profile of the Q population is probably causing the more tangential orbits observed in their $\beta(r)$ profile obtained through the IJE method. Therefore, it is important to determine whether this curvature is real and arises from intrinsic properties of typical galaxies belonging to the Q population, or if it is caused by a particular family of galaxies within this population. We found that some galaxies belonging to the Q population are classified as the brightest (rank 1) in their clusters. In general, the brightest galaxy in a cluster is also the galaxy closest to the centre of the cluster dark matter halo. Central galaxies within clusters close to equilibrium tend to present low peculiar velocities \citep[e.g.][]{Malumuth_1992_ApJ, Coziol_2009_AJ, Einasto_2012_A&A}. In addition, the peculiar velocities of their satellites are mostly influenced by the gravitational field of the central galaxy. For these reasons, all galaxies at $R/r_{200} <0.1$ have been removed from our sample (see Section~\ref{subsec:Classification_Pops}). 

    We estimate the distribution of cluster-centric distances for all rank 1 galaxies present in our sample of 588 G clusters, considering both C and O centres (see Section~\ref{subsec:Identifying_the_clusters_central_positions}). The distribution obtained considering the O centre is more elongated than that obtained with the C centre. The median value of the distributions are $0.22R/r_{200}$ for the former and $0.12R/r_{200}$ for the latter. Analysing the distribution obtained using the O centre, we verify that the cut at $R/r_{200} = 0.1$ removes several rank 1 galaxies from the sample. However, there are still many rank 1 galaxies that extend to high radial distances from the cluster centre. The correction of the centre position for each cluster, implemented in Section~\ref{subsec:Identifying_the_clusters_central_positions}, has considerably diminished this problem, with the rank 1 galaxies now being found closer to the central position of each cluster. However, the issue is not fully resolved, as there are still rank 1 galaxies with $R/r_{200} > 0.1$.

    It is important to evaluate whether these rank 1 galaxies that still remain in the sample, along with their satellites, significantly affect the MAMPOSSt results\footnote{It is important to emphasize that the sampling of central galaxies in our clusters is not necessarily complete, so these rank 1 galaxies are only indicative of the central ones and may not correspond to them exactly.}. Our main concern is related to the $M(r)$ profile, which is determined using all galaxies in the sample and is required for the $\beta(r)$ profiles estimated from IJE. In order to check the impact of such galaxies in the $M(r)$ profile estimation, we execute MAMPOSSt only for the Q population, considering all galaxies and only those with cluster-centric distances in the range $0.25 \leq R/r_{200} \leq 1.0$. The cut at $R/r_{200} \geq 0.25$ was applied to remove the majority of the rank 1 galaxies that still remain in the sample, along with their satellites. Both executions were realized as in Section~\ref{subsec:Mass_profile_Pops}, using \textit{Split} and \textit{TAND} modes of MAMPOSSt, the $r_\nu$ value from Table~\ref{tab:table_I_fit_Pops} and the redshift value for the ensemble cluster. The results obtained for the $M(r)$ profile (parameters $r_{200}$ and $r_{-2}$) are, considering the uncertainties, essentially the same. 

    We also examined the $\beta(r)$ profiles estimated by MAMPOSSt and their corresponding $\sigma_P(R)$ profiles. For the $\beta(r)$ profile of the entire sample of Q galaxies, we obtain $\beta_0 = -0.02_{-0.90}^{+0.75}$ and $\beta_\infty = 0.25_{-0.30}^{+0.24}$, showing no significant difference compared to the solution obtained using the four galaxy populations in Section~\ref{subsec:Mass_profile_Pops}. The MAMPOSSt solution for the sample of Q galaxies excluding those at $R/r_{200} < 0.25$ provided $\beta_0 = -0.43_{-0.93}^{+1.17}$ and $\beta_\infty = 0.34_{-0.34}^{+0.26}$. The lack of galaxies in the innermost regions limits our confidence in the results for this area. The $\sigma_P(R)$ profiles obtained from these $\beta(r)$ profiles are similar. However, the MAMPOSSt solution, after excluding such galaxies, yields slightly higher velocity dispersions in the internal region ($R/r_{200} \lesssim 0.45$) compared to the solution that includes all galaxies. This result is consistent with our expectations after removing the central galaxies and their satellites from the sample, namely, an increase in the velocity dispersion values. Nevertheless, the $\sigma_P(R)$ profile does not increase as much as that of IJE. This is likely due to the constraint we imposed between the $r_\nu$ and $r_\beta$ parameters in the MAMPOSSt run, which restricts the $\beta(r)$ profile and does not occur in the IJE.
        
    We therefore conclude that these rank 1 galaxies still present in the sample, together with their satellites, do not significantly affect the $M(r)$ profile obtained, and our results remain unchanged. At this point, it is important to emphasize that an efficient removal of all central galaxies in our sample is challenging. The need to impose a hard cut on $R/r_{200}$ does not guarantee the removal of all central galaxies and their satellites. The cut imposed by us at $R/r_{200} = 0.1$ has proven effective in removing most of the rank 1 galaxies and their satellites, in the sense that the remaining rank 1 galaxies do not significantly affect the $M(r)$ profile, while a sufficient number of galaxies remain in the sample to ensure the reliability of the results obtained from the dynamical analysis.
    

\section{Discussion}
\label{sec:Discussion}

    The results obtained in the previous sections show that Q galaxies are typically characterized by lower cluster-centric distances compared to SF galaxies. Additionally, Q galaxies exhibit lower velocity dispersion values at all radial distances. Similar results in the literature show that cluster passive/early-type/red galaxies are located closer to the central regions than emission-line/late-type/blue galaxies \citep[e.g.][]{Aguerri_2007_A&A, Mercurio_2021_A&A}. Moreover, early-type galaxies display steeper velocity dispersion profiles relative to late-type galaxies \citep[e.g.][]{Adami_1998_A&A_2, Biviano_2004_A&A, Biviano_2009_A&A, Cava_2017_A&A}, and red galaxies show lower velocity dispersion values compared to blue galaxies \citep[e.g.][]{Aguerri_2007_A&A}. These results suggest that quiescent galaxies have resided longer in their clusters, while SF galaxies are more recent infallers, which explains both their higher cluster-centric distances and velocity dispersion values.

    This scenario is supported by evidence indicating that as a galaxy penetrates deep into the denser cluster regions, environmental effects like ram-pressure stripping and starvation \citep[e.g.][]{Lotz_2019_MNRAS, Wright_2022_MNRAS} lead to a decrease in its gas supply, directly affecting its ability to form stars. Additionally, galaxy-galaxy and galaxy-cluster interactions reduce the kinetic energy of galaxies, thereby decreasing their peculiar velocities \citep[e.g.][]{Goto_2005_MNRAS}. This picture in which SF galaxies are recent arrivals and thus far from equilibrium, while quiescent galaxies are closer to equilibrium within their clusters, has been proposed in several works in the literature to explain a variety of observational results \citep[e.g.][]{Adami_1998_A&A_2, Biviano_2004_A&A, Mercurio_2021_A&A}. We confirm this trend by separating galaxies only according to the main gas ionization source, irrespective of the galaxy morphology or colour. 

    The results obtained for the SF population can also be explained by assuming that SF galaxies are in equilibrium within their clusters but possess higher mechanical energies given their larger cluster-centric distances and higher values of velocity dispersion. However, the dynamical modelling performed with MAMPOSSt was unable to recover the observed $\sigma_P(R)$ profile of the SF population (Fig.~\ref{fig:SigmaP_profile_Pops}). While the profiles of the Q, AGN and T populations are well described by the MAMPOSSt solutions, the SF population seems to possess an excess of kinetic energy that MAMPOSSt cannot reproduce considering equilibrium solutions, suggesting that SF galaxies as a population are not in equilibrium within their clusters. In order to better reproduce the $\sigma_P(R)$ profile of the SF population, the IJE suggests more radial orbits at all radii. However, it is worth noting that the IJE solution for $\beta(r)$ is constructed using the observed $\sigma_P(R)$ profile as input. Conversely to the SF population, MAMPOSSt reproduces the $\sigma_P(R)$ profile of the Q population very well, with the IJE solution not showing any significant improvement. These results suggest that Q galaxies are closer to equilibrium within their clusters compared to SF galaxies.


    The $\beta(r)$ profiles of Q and SF populations are consistent with inner isotropic orbits that become radial with increasingly cluster-centric distances. These profiles are in agreement with those obtained by \cite{Biviano_2013_A&A}, \cite{Biviano_2016_A&A} and \cite{Biviano_2021_A&A} for their colour-selected passive and SF galaxies. In addition, the results for the SF population are also in agreement with those obtained for emission line galaxies by \cite{Biviano_2009_A&A} and for blue galaxies by \cite{Munari_2014_A&A}. However, \cite{Biviano_2009_A&A} found that their non-emission line galaxies have fully isotropic orbits inside the virial region, while the red population of \cite{Munari_2014_A&A} is characterized by isotropic orbits in the inner regions that become somewhat tangential beyond the virial radius. On the other hand, \cite{Capasso_2019_MNRAS} found that passive galaxies are characterized by isotropic orbits in the inner regions and increasingly radial at larger radii, in agreement with our findings for the Q population. Additionally, the $\beta(r)$ profiles of high- and low-mass passive galaxies found by \cite{Annunziatella_2016_A&A} marginally agree with our profile for the Q galaxies. However, their external anisotropies for both populations are very high ($\beta > 0.55$ at $r_{200}$) compared to what we find for the Q population; also, their internal anisotropy for the low-mass sample indicates tangential orbits. In \cite{Mamon_2019_A&A} the authors find that the kinematics of elliptical and lenticular galaxies is consistent with isotropy everywhere, although mildly radial orbits at $r_{200}$ are also acceptable for both populations. In this case, the lenticular population ($\beta_{\rm{S0}} \simeq 0.31 \pm 0.17$) exhibits more radial orbits at $r_{200}$ than the elliptical population ($\beta_{\rm{E}} \simeq 0.19 \pm 0.25$). In contrast, spiral galaxies are characterized by isotropic orbits in the inner regions that become more radial ($\beta \simeq 0.45 \pm 0.08$) at $r_{200}$. Therefore, our results for the Q galaxies are similar to those of the elliptical and lenticular populations, while the SF orbits more closely resemble those of the spiral galaxies. 
    
    We also find that SF galaxies are characterized by more radial orbits compared to Q galaxies, especially in the clusters outskirts (Fig.~\ref{fig:Beta_profile_Pops}). This result, although marginally significant, is consistent with several studies in the literature, which have already reported that star-forming/late-type/blue galaxies tend to be characterized by more radial orbits than quiescent/early-type/red galaxies, both in observations \citep[e.g.][]{Biviano_2004_A&A, Biviano_2009_A&A, Munari_2014_A&A, Mamon_2019_A&A, Biviano_2021_A&A} and simulations \citep[e.g.][]{Lotz_2019_MNRAS}. Such results can be explained by considering the physical processes that govern the evolution of galaxy clusters. Quiescent galaxies, having resided longer in the cluster environment, lose their orbital information during the initial phase of fast collapse of these systems due to an efficient dynamical relaxation process. In contrast, star-forming systems are accreted into the cluster during a later phase, where the cluster grows slowly by accreting matter from its surroundings. As a result, these galaxies still retain the orbital information from their infall epoch \citep[e.g.][]{Lapi_2011_ApJ}. Despite this, there are also works that found the opposite trend, with red galaxies exhibiting more radial orbits than blue galaxies \citep[e.g.][]{Aguerri_2017_MNRAS, Mercurio_2021_A&A}.
    
    In our classification scheme, SF galaxies are characterized by gas ionization due primarily by young, massive stars within the SDSS fibre \citep[2 arcsec,][]{Eisenstein_2011_AJ} at the time of observation, without any constraints regarding the colour or morphology. Hence, the SF population includes extreme objects like starburst galaxies and red early-type galaxies that underwent `rejuvenation' by minor merging or cold gas infall. Q galaxies, on the other hand, present only weak, if any, emission lines within the SDSS fibre. As a result, nearby blue late-type galaxies, where the gas ionized mainly by stars is distributed across the disk and does not contribute to the ionization within the SDSS fibre, can be classified as Q galaxies. Consequently, the Q population is not equivalent to a population of elliptical galaxies, or even to a population of red galaxies. Therefore, it is worth noting that the trends observed between passive/early-type/red galaxies and emission line/late-type/blue galaxies, regarding their spatial distribution \citep[e.g.][]{Aguerri_2007_A&A, Mercurio_2021_A&A, Cava_2017_A&A}, velocity dispersion \citep[e.g.][]{Adami_1998_A&A_2, Biviano_2004_A&A, Biviano_2009_A&A, Cava_2017_A&A} and orbital properties \citep[e.g.][]{Biviano_2009_A&A, Munari_2014_A&A, Biviano_2013_A&A, Biviano_2016_A&A, Mamon_2019_A&A, Biviano_2021_A&A}, remain the same when we look exclusively at the gas ionization source contained in the spectroscopic fibre.
 
    The SF population is thought to be composed of recent infallers \citep[e.g.][]{Lotz_2019_MNRAS}, which would lead us to expect these galaxies to display radial orbits at all radii. Conversely, Q galaxies, having likely resided longer in the cluster environment, are expected to be characterized by more isotropic orbits \citep[e.g.][]{Biviano_1997_A&A, Biviano_2009_A&A}. However, we observe that both populations exhibit $\beta(r)$ profiles that rise from nearly isotropic in the inner regions to increasingly radial with radius. The outer anisotropy of the SF population likely arises due to the presence of newly arrived galaxies that still retain their more radial orbits from the infall period. Since these galaxies have not spent much time in the cluster environment, they still retain their gas reservoirs, allowing them to continue forming stars and thus appear as SF galaxies. As these systems move towards the denser central regions, environmental mechanisms start to act. Some of these processes, such as ram-pressure \citep[e.g.][]{Gunn_1972_ApJ}, are responsible for starting the removal of the gas component of these galaxies \citep[e.g.][]{Farouki_1980_ApJ, Abadi_1999_MNRAS, Boselli_2006_PASP}, while processes like dynamical friction and tidal braking can lead to the isotropization of the orbits. Since ram pressure is more effective at suppressing star formation in objects with more radial orbits or lower stellar mass \citep[e.g.][]{Vollmer_2001_ApJ, Jaffe_2018_MNRAS, Lotz_2019_MNRAS}, these will abandon the SF population and become quiescent galaxies. Consequently, some of these recently quenched galaxies will be responsible for the more radial orbits observed for the Q population at larger cluster-centric distances. Additionally, galaxies that suffered pre-processing at the group environment and quenched their star formation \citep[e.g.][]{Fujita_2004_PASJ, Bianconi_2018_MNRAS,Lopes_2024_MNRAS}, will also contribute with higher orbital anisotropies to the Q population in the cluster outskirts. On the other hand, galaxies that remain capable of retaining their gas supply and keep forming stars will be characterized by less radial orbits as we move to the cluster central regions. This could explain the trend towards less radial orbits at smaller radii observed for the SF population. An interesting result that corroborates this idea was obtained by \cite{Lotz_2019_MNRAS}, who analysed which characteristics distinguish satellite galaxies that remain capable of forming stars for longer periods compared to the total population, in the \emph{Magneticum Pathfinder} simulations. Their results show that among high stellar mass satellite galaxies, those with the greatest stellar masses are more likely to survive. In contrast, the low stellar mass satellites that survive are those with more tangential orbits. Additionally, quiescent galaxies that reside longer in the cluster environment but recently accreted gas by some mechanism such as gas-rich mergers \citep[e.g.][]{Mihos_1996_ApJ, Springel_2005_MNRAS, Hopkins_2006_ApJS} and are now forming stars will appear to us as SF galaxies. These objects are characterized by more isotropic orbits and tend to reside deeper in the cluster environment \citep[e.g.][]{Biviano_2009_A&A, Lotz_2019_MNRAS}. Thus, they will contribute to the lower anisotropies observed for the SF population at lower cluster-centric distances.

    Galaxies belonging to the T population tend to be less spatially concentrated towards the centres of the clusters than those of the Q population, but are still more concentrated than those of the SF population (Tab.~\ref{tab:table_I_fit_Pops}). The velocity dispersion values of the T population also tend to fall between those of the Q and SF populations (see Fig.~\ref{fig:SigmaP_profile_Pops}). These results suggest that, if we interpret Q systems as galaxies closer to equilibrium within the cluster environment, while the SF population consists of recent infallers \citep[e.g.][]{Lotz_2019_MNRAS},  T galaxies are likely in an intermediate dynamical stage between these two populations. In contrast to the T population, we find that AGN host galaxies are as concentrated around the cluster centres as Q galaxies. Nevertheless, the velocity dispersion values of the AGN population tend to be higher than those of the Q population, being more similar to those of SF galaxies, suggesting that AGN hosts are less dynamically evolved than Q systems in the cluster environments. The results obtained for the T and AGN populations in this paper are similar to those reported by \cite{Lopes_2017_MNRAS}. By analysing the distributions of transition, AGN, passive and SF galaxies in the PPS, the authors identified that transition and AGN galaxies exhibit distributions distinct from each other and from the passive and SF populations. Additionally, based on the PPS distribution, they concluded that AGN and T galaxies are composed of a mix of early and recent infallers, while the passive and SF populations are mainly composed of early and recent infallers, respectively. 
    The dynamical results obtained for the T and AGN populations support the scenario in which these are in an intermediate dynamical stage between those of the Q and SF populations. The MAMPOSSt and IJE solutions for the $\beta(r)$ profiles of both populations are consistent with inner isotropic orbits that become more elongated with increasingly cluster-centric distances. Additionally, the observed $\sigma_P(R)$ profiles of both populations are successfully reproduced by the MAMPOSSt and IJE solutions for the $\beta(r)$ profile. To the best of our knowledge, this is the first determination of the orbital profiles of T galaxies and AGN hosts selected according to optical emission line diagnostic diagrams.

    Transition galaxies are objects that possess a reservoir of ionized gas and exhibit mixed ionization \citep[SF+AGN, e.g.][]{Ho_1993_ApJ, Kauffmann_2003_MNRAS, Kewley_2006_MNRAS}. Nevertheless, the AGN population also includes objects that are actively forming stars, and therefore both populations are at least partially dependent on a cold gas reservoir. As SF and late-type galaxies in general are thought to be recent infallers \citep[e.g.][]{Biviano_2004_A&A, Lotz_2019_MNRAS, Mamon_2019_A&A} that eventually settle in an equilibrium configuration as they lose their cold gas reservoirs, the similarities between the kinematical properties of SF and T/AGN populations would naturally arise in a scenario where star formation and optical nuclear activity are mostly dependent on a common cold gas supply. Additionally, the AGN and T populations include galaxies in which star formation and/or nuclear activity have been triggered recently due to environmental mechanisms. In particular, if mergers are the main triggering mechanism of SF and/or nuclear activity, causing the cold gas to lose angular momentum and fall towards the galaxy's central region \citep[e.g.][]{Noguchi_1988_A&A, Hernquist_1995_ApJ, Mihos_1996_ApJ, Gao_2020_A&A}, they are expected to occur with higher efficiency in the cluster outskirts during infall. In fact, since mergers require lower relative velocities between the involved galaxies, they are commonly invoked to explain the higher AGN fractions found in the outskirts of galaxy clusters \citep[e.g.][]{Lopes_2017_MNRAS, Koulouridis_2024_A&A}. Objects recently accreted into the cluster environment that triggered SF and/or AGN activity during their infall phase will tend to contribute to the higher velocity dispersion values and the more radial orbits observed in the profiles of the T and AGN populations. On the other hand, environmental processes, such as galaxy harassment, can also trigger SF and/or nuclear activity in those galaxies that reside longer within the cluster environment \citep[e.g.][]{Moore_1996_Natur, Moore_1998_ApJ, Byrd_1990_ApJ}. Additionally, the gas compression caused by ram-pressure stripping is known to induce short periods of enhanced star formation \citep[e.g.][]{Bekki_2003_ApJL, Kapferer_2009_A&A, Kronberger_2008_A&A, Tonnesen_2009_ApJ}. Moreover, recent studies have reported a higher fraction of nuclear activity in galaxies undergoing a stronger ram-pressure stripping effect \citep[e.g.][]{Poggianti_2017_Natur, Poggianti_2021_IAUS, Peluso_2022_ApJ}, known as `jellyfish' galaxies \citep[e.g.][]{Ebeling_2014_ApJL}, compared to non-ram pressure stripped galaxies, thus indicating a possible correlation between ram-pressure and AGN. Because the galaxies that reside deeper in the cluster potential are expected to be characterized by more isotropic orbits, these objects could explain the more concentrated distributions exhibited by the AGN and T populations, as well as the gradient towards isotropic orbits at smaller radii in their $\beta(r)$ profiles.

    Since optical AGNs are commonly found in more massive galaxies \citep[e.g.][]{Kauffmann_2003_MNRAS, Pimbblet_2013_MNRAS}, AGN and T galaxies will typically be more massive than those in the SF population. As a result, they will be able to retain their gas reservoirs for longer periods compared to typical star-forming galaxies while falling towards the denser central regions, given that ram-pressure is more effective on lower-mass objects \citep[e.g.][]{Lotz_2019_MNRAS}. Consequently, T and AGN galaxies could retain their respective ionization signatures for longer periods than typical SF galaxies. On the other hand, AGNs hosts are expected to suffer stronger dynamical friction compared to typical SF galaxies due to their higher stellar masses \citep[e.g.][]{Merritt_1983_ApJ, Lotz_2019_MNRAS}. We thus expect that AGN and T galaxies will lose their dynamical information from the infall phase more quickly than typical SF galaxies. This helps to explain the more concentrated spatial distribution exhibited by the AGN and T populations compared to the SF population and the tendency of T galaxies to exhibit lower velocity dispersion values than SF galaxies. In contrast, although the AGN and SF populations appear to have similar velocity dispersion values (Fig.~\ref{fig:SigmaP_profile_Pops}), it is difficult to compare both profiles due to the higher uncertainties in the $\sigma_P(R)$ profile of the AGN population. One interesting result we find is that galaxies that only host an optical AGN tend to be more concentrated than galaxies that display AGN+SF ionization, suggesting that star formation quenching is faster than the suppression of optical nuclear activity. In this context, an indication that nuclear activity and star formation are not affected by the same processes to the same degree in clusters was obtained by \cite{Rihtarsic_2024_A&A} analysing the X-ray AGN fraction in massive galaxies using the \emph{Magneticum Pathfinder} simulations. Consequently, we are able to observe galaxies with AGN activity deeper into the cluster potential, while star formation has already been suppressed. In particular, this also implies that the AGN population has been accreted in earlier epochs into the cluster environment compared to T galaxies.


    The results presented in this work suggest a dynamical evolutionary scenario for the galaxies belonging to the Q, AGN, T, and SF populations. The SF population is composed of a large fraction of recent arrivals to the cluster environment that still retain the properties from their infall epoch. As these SF galaxies gradually penetrate into denser regions, environmental effects remove their gas, leading to the quenching of star formation. Simultaneously, encounters with other galaxies isotropize their velocities, reducing their velocity dispersions. By the end of this process, accreted SF galaxies will evolve into a more virialized Q population, with quenched star formation, lower velocity dispersions, and concentrated in the inner regions of clusters. In this scenario, the results obtained for the T and AGN galaxies are consistent with these objects being situated in an intermediate dynamical state between those of the Q and SF populations, with AGN hosts marginally more virialized than the T population.
    


\section{Conclusions}
\label{sec:Conclusions}

    In this work, we analysed the kinematical and dynamical properties of four cluster galaxy populations classified according to the dominant source of gas ionization, namely: star-forming (SF), optically active galactic nuclei (AGN), transition objects (T, SF+AGN ionization) and quiescent (Q). Our sample is composed of 8\,892 galaxies belonging to 336 relaxed galaxy clusters ($z<0.2$).

    Our main results are summarized below:

    \begin{itemize}
        
        \item [(i)] The Q population is the closest to equilibrium. Quiescent galaxies are the most concentrated, exhibit the lowest velocity dispersion values, and their observed $\sigma_P(R)$ profile is accurately recovered by MAMPOSSt under equilibrium assumptions.
        
        \item [(ii)] The SF population shows the clearest signs of non-equilibrium: the highest typical cluster-centric distances, higher velocity dispersion values, and an excess of velocity dispersions relative to the MAMPOSSt equilibrium solution.  
    
        \item [(iii)] Transition galaxies are in a dynamical state intermediate between those of the Q and SF populations. The T population exhibits typical cluster-centric distances and velocity dispersion values that are intermediate between those of the Q and SF populations. Additionally, MAMPOSSt was able to satisfactorily recover their observed $\sigma_P(R)$ profile.

        \item [(iv)] The AGN population consists of galaxies as concentrated as those in the Q population, but with velocity dispersion values more similar to those of SF galaxies. Furthermore, their $\sigma_P(R)$ profile was satisfactorily recovered by MAMPOSSt under equilibrium assumptions. This likely indicates that AGN galaxies are even more virialized than those in the T population but still less relaxed than those in the Q population. 

        \item [(v)] The $\beta(r)$ profiles of all populations are consistent with isotropic orbits in the inner regions that become increasingly radial with increasing cluster-centric distances.

        \item [(vi)] The SF, AGN, and T galaxies appear to be characterized by more radial orbits at $r_{200}$ than Q galaxies. However, this conclusion is only marginally significant due to the high uncertainties in the $\beta(r)$ profiles.
        
        \end{itemize}

    We interpret these results within a scenario in which the SF population consists of recent infallers that still retain kinematical and dynamical information from their infall epoch. As these galaxies penetrate deeper into the denser regions of clusters, environmental effects remove their gas, leading to the quenching of star formation while encounters with other galaxies reduce their velocity dispersions. These SF galaxies evolve towards a more virialized Q population, becoming more concentrated in the inner regions and exhibiting lower velocity dispersion values. The T and AGN populations consist of galaxies that are in an intermediate dynamical stage between those of the Q and SF populations, likely evolving towards quiescence.
    

\section*{Acknowledgements}

The authors acknowledge support from the Brazilian research agencies Conselho Nacional de Desenvolvimento Científico e Tecnológico (CNPq) and Fundação de Amparo à pesquisa do Estado do Rio Grande do Sul (FAPERGS). GAV thanks for the financial support from Coordenação de Aperfeiçoamento de Pessoal de Nível Superior -- Brasil (CAPES) -- Finance Code 001.

Funding for SDSS-III has been provided by the Alfred P. Sloan Foundation, the Participating Institutions, the National Science Foundation, and the U.S. Department of Energy Office of Science. The SDSS-III web site is \url{http://www.sdss3.org/}.

SDSS-III is managed by the Astrophysical Research Consortium for the Participating Institutions of the SDSS-III Collaboration including the University of Arizona, the Brazilian Participation Group, Brookhaven National Laboratory, Carnegie Mellon University, University of Florida, the French Participation Group, the German Participation Group, Harvard University, the Instituto de Astrofisica de Canarias, the Michigan State/Notre Dame/JINA Participation Group, Johns Hopkins University, Lawrence Berkeley National Laboratory, Max Planck Institute for Astrophysics, Max Planck Institute for Extraterrestrial Physics, New Mexico State University, New York University, Ohio State University, Pennsylvania State University, University of Portsmouth, Princeton University, the Spanish Participation Group, University of Tokyo, University of Utah, Vanderbilt University, University of Virginia, University of Washington, and Yale University.

\section*{Data Availability}

No new data have been produced in this work.



\bibliographystyle{mnras}
\bibliography{bibliography_valk} 



\appendix


\bsp	
\label{lastpage}
\end{document}